# GJ 367b: A dense ultra-short period sub-Earth planet transiting a nearby red dwarf star

**Authors:** Kristine W. F. Lam[1,2*†], Szilárd Csizmadia[2†], Nicola Astudillo-Defru[3], Xavier Bonfils[4], Davide Gandolfi[5], Sebastiano Padovan[2,46], Massimiliano Esposito[6], Coel Hellier[7], Teruyuki Hirano[8], John Livingston[9], Felipe Murgas[10,11], Alexis M. S. Smith[2], Karen A. Collins[12], Savita Mathur[10,11], Rafael A. Garcia[13,14], Steve B. Howell[21], Nuno C. Santos[15,16], Fei Dai[27], George R. Ricker[17], Roland Vanderspek[17], David W. Latham[12], Sara Seager[17,18,19], Joshua N. Winn[20], Jon M. Jenkins[21], Simon Albrecht[22], Jose M. Almenara[4], Etienne Artigau[4], Oscar Barragán[23], François Bouchy[24], Juan Cabrera[4], David Charbonneau[12], Priyanka Chaturvedi[6], Alexander Chaushev[1], Jessie L. Christiansen[25], William D. Cochran[26], José R. De Meideiros[28], Xavier Delfosse[4], Rodrigo F. Díaz[29], René Doyon[30], Philipp Eigmüller[2], Pedro Figueira[31,15], Thierry Forveille[4], Malcolm Fridlund[32,33], Guillaume Gaisné[4], Elisa Goffo[5,6], Iskra Georgieva[32], Sascha Grziwa[34], Eike Guenther[6], Artie P. Hatzes[6], Marshall C. Johnson[35], Petr Kabáth[36], Emil Knudstrup[22], Judith Korth[34,49], Pablo Lewin[37], Jack J. Lissauer[21,48], Christophe Lovis[24], Rafael Luque[10,11], Claudio Melo[31], Edward H. Morgan[17], Robert Morris[21,38], Michel Mayor[24], Norio Narita[39,40,41,10], Hannah L. M. Osborne[42], Enric Palle[10,11], Francesco Pepe[24], Carina M. Persson[32], Samuel N. Quinn[12], Heike Rauer[2,1,43], Seth Redfield[44], Joshua E. Schlieder[45], Damien Ségransan[24], Luisa M. Serrano[5], Jeffrey C. Smith[21,38], Ján Šubjak[36,47], Joseph D. Twicken[21,38], Stéphane Udry[24], Vincent Van Eylen[42], Michael Vezie[17]

**Affiliations:**
[1] Centre for Astronomy and Astrophysics, Technical University Berlin, 10585 Berlin, Germany
[2] Institute of Planetary Research, German Aerospace Center, 12489 Berlin, Germany
[3] Departamento de Matemática y Física Aplicadas, Universidad Católica de la Santísima Concepción, Concepción, Chile
[4] Université Grenoble Alpes, Centre national de la recherche scientifique, Institut de Planétologie et d'Astrophysique de Grenoble, F-38000 Grenoble, France
[5] Dipartimento di Fisica, Università degli Studi di Torino, I-10125, Torino, Italy
[6] Thüringer Landessternwarte Tautenburg, D-07778 Tautenburg, Germany
[7] Astrophysics Group, Keele University, Staffordshire, ST5 5BG, UK
[8] Department of Earth and Planetary Sciences, Tokyo Institute of Technology, Tokyo, Japan
[9] Department of Astronomy, University of Tokyo, Tokyo 113-0033, Japan
[10] Instituto de Astrofísica de Canarias, 38205 La Laguna, Tenerife, Spain
[11] Departamento de Astrofísica, Universidad de La Laguna, 38206, La Laguna, Tenerife, Spain
[12] Center for Astrophysics, Harvard & Smithsonian, Cambridge, MA 02138, USA
[13] Institut de Recherche sur les Lois Fondamentales de l'Universe, Commissariat à l'Énergie Atomique et aux énergies alternatives, Université Paris-Saclay, F-91191 Gif-sur-Yvette, France
[14] Astrophysique, Instrumentation et modélisation, Commissariat à l'Énergie Atomique et aux énergies alternatives, Centre National de la recherche scientifique, Université Paris-Saclay, Université Paris Diderot, Sorbonne Paris Cité, F-91191 Gif-sur-Yvette, France

[15] Instituto de Astrofísica e Ciênciasdo Espaço, Universidade do Porto, Centro de Astrofísica da Universidade do Porto, 4150-762 Porto, Portugal

[16] Departamento de Física e Astronomia, Faculdade de Ciências, Universidade do Porto, 4169-007 Porto, Portugal

[17] Department of Physics and Kavli Institute for Astrophysics and Space Research, Massachusetts Institute of Technology, Cambridge, MA 02139, USA

[18] Department of Earth, Atmospheric and Planetary Sciences, Massachusetts Institute of Technology, Cambridge, MA 02139, USA

[19] Department of Aeronautics and Astronautics, Massachusetts Institute of Technology, Cambridge, MA 02139, USA

[20] Department of Astrophysical Sciences, Princeton University, Princeton, NJ 08544, USA

[21] NASA Ames Research Center, Moffett Field, CA, 94035, USA

[22] Stellar Astrophysics Centre, Department of Physics and Astronomy, Aarhus University, DK-8000 Aarhus C, Denmark

[23] Sub-department of Astrophysics, Department of Physics, University of Oxford, Oxford OX1 3RH, UK

[24] Geneva Observatory, University of Geneva, 1290 Versoix, Switzerland

[25] Infrared Processing and Analysis Center, Caltech, Pasadena CA 91125, USA

[26] Center for Planetary Systems Habitability, and McDonald Observatory, The University of Texas, Austin Texas, USA

[27] Division of Geological and Planetary Sciences, California Institute of Technology, Pasadena, CA, 91125, USA

[28] Departamento de Física, Universidade Federal do Rio Grande do Norte, 59072-970 Natal, RN, Brazil

[29] International Center for Advanced Studies and Instituto de Ciencias Físicas (Consejo Nacional de Investigaciones Científicas y Técnicas), Escuela de Ciencia y Tecnología - Universidad Nacional de San Martín, Campus Miguelete, Buenos Aires, Argentina

[30] Institut de Recherche sur les Exoplantes, Dpartement de Physique, Universit de Montral, Montral, QC, H3C 3J7, Canada

[31] European Southern Observatory, Vitacura, Santiago, Chile

[32] Department of Space, Earth and Environment, Chalmers University of Technology, Onsala Space Observatory, 439 92 Onsala, Sweden

[33] Leiden Observatory, University of Leiden, PO Box 9513, 2300 RA, Leiden, The Netherlands

[34] Rheinisches Institut für Umweltforschung an der Universität zu Köln, D-50931 Köln Germany

[35] Las Cumbres Observatory, Goleta, CA 93117, USA

[36] Astronomical Institute, Czech Academy of Sciences, 25165, Ondřejov, Czech Republic

[37] The Maury Lewin Astronomical Observatory, Glendora, California 91741. USA

[38] Search for Extraterrestrial Intelligence Institute, Mountain View, CA 94043, USA

[39] Komaba Institute for Science, The University of Tokyo, Tokyo 153-8902, Japan

[40] Japan Science and Technology Agency, Precursory Research for Embryonic Science and Technology, Tokyo 153-8902, Japan

[41] Astrobiology Center, Tokyo 181-8588, Japan

[42] Mullard Space Science Laboratory, University College London, Dorking, Surrey, RH5 6NT, UK

[43] Institute of Geological Sciences, Freie Universität Berlin, D-12249 Berlin

[44] Astronomy Department and Van Vleck Observatory, Wesleyan University, Middletown, CT 06459, USA

[45] NASA Goddard Space Flight Center, Greenbelt, MD 20771, USA

[46] WorkGroup Solutions GmbH at European Organisation for the Exploitation of Meteorological Satellites, 64295 Darmstadt, Germany.

[47] Astronomical Institute of Charles University, 180 00 Prague, Czech Republic

[48] Geological Sciences Department, Stanford University, CA 94305-2115, USA

[49] Department of Space, Earth and Environment, Astronomy and Plasma Physics, Chalmers University of Technology, 412 96 Gothenburg, Sweden

*Corresponding author. Email: Kristine.Lam@dlr.de

†These authors contributed equally to this work.

**Abstract:**

Ultra-short-period (USP) exoplanets have orbital periods shorter than one day. Precise masses and radii of USPs could provide constraints on their unknown formation and evolution processes. We report the detection and characterization of the USP planet GJ 367b using high precision photometry and radial velocity observations. GJ 367b orbits a bright (*V*-band magnitude = 10.2), nearby, red (M-type) dwarf star every 7.7 hours. GJ 367b has a radius of $0.718 \pm 0.054$ Earth-radii, a mass of $0.546 \pm 0.078$ Earth-masses, making it a sub-Earth. The corresponding bulk density is $8.106 \pm 2.165$ g cm$^{-3}$, close to that of iron. An interior structure model predicts the planet has an iron core radius fraction of $86 \pm 5\%$, similar to Mercury's interior.

**Main Text:**

Red dwarf stars of spectral type M (M dwarfs) are cool stars with effective temperatures ($T_{\mathrm{eff}}$) below ~4000 K *(1)*. They have masses and radii less than around ~60% of the Sun's and are the most abundant type of stars in the solar neighborhood *(2–4)*. It has been estimated that M dwarfs host an average of $2.5 \pm 0.2$ small planets [planet radius $R_{\mathrm{p}} < 4$ Earth-radius ($R_{\oplus}$)] with periods less than 100 days *(5)*. Due to the small stellar radius, the transit signal produced by a planet orbiting an M dwarf is larger than a planet of the same size orbiting a solar-type star (G dwarf). The radial velocity (RV) signal induced by a planet is also larger for an M dwarf host than for a G dwarf, due to the lower stellar mass. M dwarfs therefore provide an opportunity to search for exoplanets with small radius and low mass. However, young M dwarfs often have high stellar activity, inducing noise in the RV observations *(6)*. RV analysis can be complicated even for old, inactive M dwarfs because their slow rotation periods have harmonics in the range of periods where small planets are sought *(7)*.

GJ 367 (also catalogued as TOI-731) is an M dwarf at 9.41 parsecs (pc) from the Sun *(8)* with a brightness of 10.153 magnitudes in the optical *V*-band and 5.78 magnitudes in the infrared *K*-band. We observed this star with the High Accuracy Radial velocity Planet Searcher (HARPS) spectrograph *(9)* and determined its stellar properties. GJ 367 has an effective temperature of $T_{eff}$ = 3519 ± 70 K, stellar mass $M_s$ = 0.454 ± 0.011 solar masses ($M_\odot$), stellar radius $R_s$ = 0.457 ± 0.013 solar radii ($R_\odot$), and stellar luminosity $L_s$ = 0.0288 ± 00029 solar luminosities ($L_\odot$) *(9)* (Table 1).

The Transiting Exoplanet Survey Satellite (TESS) *(10)* observed GJ 367 during sector 9 of its survey. TESS acquired optical photometry at 2-minute cadence for 27 days from 2019 February 28 to 2019 March 26. The light curve (brightness as a function of time) was extracted using the Science Processing Operations Center (SPOC) pipeline *(11)*. This revealed a planet candidate with orbital period 0.32 days and radius 0.75 $R_\oplus$, which was designated TOI-731.01 by the TESS Science Office based on the SPOC transit search and data validation results. We also searched for transit signals using the Détection Spécialisée de Transits (DST) algorithm *(12)*, which indicated a transit-like signal every 0.32 days and a transit depth of around 0.03%, corresponding to the transit of a sub-Earth size planet (Figure. 1).

We performed several tests to ensure the candidate was not a false positive. Comparison of photometric data using varying aperture sizes showed no correlation between the aperture size and transit depth, indicating that the transit signal is not from another source blended with GJ 367. We performed follow-up ground-based photometry, finding no contamination from eclipsing binaries up to 2.5 arcsec from the target star *(9)*. The follow-up photometry shows that nearby stellar sources contribute 9.5 ± 1.2% of the flux within the TESS optimal aperture. This contamination reduces the transit depth, causing analysis of the TESS light curve to underestimate the planet radius by about 5% *(9)*. We account for this dilution factor in a transit model to obtain the true planet radius (Table 1). The density of the host star, $\rho_s$, was derived from the transit light curve (9), finding 7.64 ± 3.51 g cm⁻³, which is consistent with the value $\rho_s = 6.71^{+0.61}_{-0.55}$ g cm⁻³ determined from the spectral analysis discussed above *(9)*.

In a further test, we performed a frequency analysis of the HARPS RV measurements and stellar activity indicators *(9)*. The periodogram of the RVs has a peak at orbital frequency (*f*) of 3.103 d⁻¹ (*P* = 0.322 days) that has no counterpart in the periodograms of the activity indicators (Figure S4), consistent with a planetary origin. A further 45-day signal is present in the RV periodogram and in the activity indicators. Our analysis of archival photometry from the Wide Angle Search for Planets (WASP) survey indicates a stellar rotational period of 48 ± 2 days *(9)*. GJ 367's Ca II activity index is $\log R'_{HK} = -5.214 \pm 0.074$, which corresponds to an estimated stellar rotation period of 58.0 ± 6.9 days *(9)*. This indicates that the 45-day signal likely originates from active regions on the stellar surface. We conclude that the 0.322-day period is due to a USP planet, GJ 367b.

Using a priori information on the host star properties from our spectral analysis, we derived the physical properties of the GJ 367 system using a Bayesian Markov-chain Monte Carlo (MCMC) code, TRANSIT AND LIGHT CURVE MODELLER *(13)*, to model the transit photometry and RV data simultaneously *(9)*. Table 1 reports the physical properties of the planetary system from this analysis. The transit depth of 212 ± 42 ppm and RV semi-amplitude $K$ = 79.8 ± 11.0 cm s$^{-1}$ correspond to a planetary radius of 0.718 ± 0.054 $R_\oplus$ and a planetary mass of 0.546 ± 0.078 Earth-masses ($M_\oplus$). Figure 1 shows the phase-folded light curve and RV measurements of GJ 367, along with the corresponding best-fitting transit and RV models. We find that GJ 367b is a sub-Earth planet with a high expected signal-to-noise metric for emission spectroscopy (see Supplementary Text). The planet receives high stellar irradiation due to its close proximity to the host star, about 576 times the incident flux on Earth. This corresponds to a dayside temperature of 1745 ± 43 K (assuming zero Bond albedo), high enough to evaporate any primordial atmosphere *(14–16)* and begin to melt or vaporize any silicates or metallic iron *(17)*.

The measured mass and radius of GJ 367b imply a bulk density of 8.106 ± 2.165 g cm$^{-3}$, which we compare to the other USP planets in Figure 3A. The bulk composition of a planet can be estimated from theoretical mass-radius relations *(18–21)*. Figure 2 shows the mass and radius distribution of small planets ($R_p$ below 2 $R_\oplus$) along with theoretical predictions for rocky planets *(21, 22)*. GJ 367b has a mass and radius consistent with an interior dominated by an iron core. This is similar to two larger USP planets, K2-229b *(23)* and K2-141b *(24, 25)*, which have enhanced iron fractions (Figure 3). Other known planets with similar sizes to GJ 367b, such as Kepler-138 b *(26, 27)* and TRAPPIST-1 d *(28, 29)*, have lower densities, longer orbital periods and are exposed to lower stellar irradiation, so may be less susceptible to loss of an atmosphere *(14)*.

We used a neural network to investigate possible interior structures of GJ 367b *(9)*. At the best-fitting density, this indicates that GJ 367b is predominantly made of iron (Figure 3B): 86 ± 5% iron core (by radius), <10% water ice and/or H & He gas envelope, and the remainder as silicate mantle. This composition is similar to that of Mercury, which the neural network predicts would have an iron core radius fraction of $81 \pm 4\%$ *(9)*. This is consistent with the measured Mercury core radius of $2030 \pm 37$ km *(30)*, which corresponds to a core radius fraction of $83 \pm 2\%$. For comparison, the interior structures of Mercury and other terrestrial planets predicted by our analysis are shown in Figure S8.

**References and Notes:**

1. C. Cifuentes, J. A. Caballero, M. Cortés-Contreras, D. Montes, F. J. Abellán, R. Dorda, et al., CARMENES input catalogue of M dwarfs. V. Luminosities, colours, and spectral energy distributions, *Astron. Astrophys.* **642**, A115 (2020).

2. E. E. Salpeter, The Luminosity Function and Stellar Evolution., *Astrophys. J.* **121**, 161 (1955).

3. T. J. Henry, W. Jao, J. P. Subasavage, T. D. Beaulieu, P. A. Ianna, E. Costa, et al., The Solar Neighborhood. XVII. Parallax Results from the CTIOPI 0.9 m Program: 20 New Members of the RECONS 10 Parsec Sample, *Astron. J.* **132**, 2360 (2006).

4. G. Chabrier, Galactic Stellar and Substellar Initial Mass Function, *Publ. Astron. Soc. Pac.* **115**, 763 (2003).

5. C. D. Dressing, D. Charbonneau, The Occurrence of Potentially Habitable Planets Orbiting M Dwarfs Estimated from the Full Kepler Dataset and an Empirical Measurement of the Detection Sensitivity, *Astrophys. J.* **807**, 45 (2015).

6. A. Suárez Mascareño, J. P. Faria, P. Figueira, C. Lovis, M. Damasso, J. I. González Hernández, et al., Revisiting Proxima with ESPRESSO, *Astron. Astrophys.* **639**, A77

(2020).

7.  S. H. Saar, R. A. Donahue, Activity-Related Radial Velocity Variation in Cool Stars, *Astrophys. J.* **485**, 319 (1997).

8.  Gaia Collaboration, A. G. A. Brown, A. Vallenari, T. Prusti, J. H. J. de Bruijne, C. Babusiaux, et al., Gaia Data Release 2. Summary of the contents and survey properties, *Astron. Astrophys.* **616**, A1 (2018).

9.  Materials and methods are available as supplementary materials

10. G. R. Ricker, J. N. Winn, R. Vanderspeck, D. W. Latham, G. Á. Bakos, J. L. Bean, et al., *Space Telescopes and Instrumentation 2016: Optical, Infrared, and Millimeter Wave* (2016), vol. 9904 of *Society of Photo-Optical Instrumentation Engineers (SPIE) Conference Series*, p. 99042B.

11. J. M. Jenkins, H. Chandrasekaran, S. D. McCauliff, D. A. Caldwell, P. Tenenbaum, J. Li, T. C. Klaus, *et al.*, *Software and Cyberinfrastructure for Astronomy IV* (2016), vol. 9913 of *Society of Photo-Optical Instrumentation Engineers (SPIE) Conference Series*, p. 99133E.

12. J. Cabrera, S. Csizmadia, A. Erikson, H. Rauer, S. Kirste, A study of the performance of the transit detection tool DST in space-based surveys. Application of the CoRoT pipeline to Kepler data, *Astron. Astrophys.* **548**, A44 (2012).

13. S. Csizmadia, The Transit and Light Curve Modeller, *MNRAS* **496**, 4442 (2020).

14. J. E. Owen, Y. Wu, The Evaporation Valley in the Kepler Planets, *Astrophys. J.* **847**, 29 (2017).

15. E. D. Lopez, J. J. Fortney, Understanding the Mass-Radius Relation for Sub-neptunes: Radius as a Proxy for Composition, *Astrophys. J.* **792**, 1 (2014).

16. J. E. Owen, Y. Wu, Kepler Planets: A Tale of Evaporation, *Astrophys. J.* **775**, 105 (2013).

17. A. Léger, O. Grasset, B. Fegley, F. Codron, A. F. Albarede, et al., The extreme physical properties of the CoRoT-7b super-Earth, *Icarus* **213**, 1 (2011).

18. D. Valencia, D. D. Sasselov, R. J. O'Connell, Detailed Models of Super-Earths: How Well Can We Infer Bulk Properties?, *Astrophys. J.* **665**, 1413 (2007).

19. L. Zeng, D. Sasselov, A Detailed Model Grid for Solid Planets from 0.1 through 100 Earth Masses, *Publ. Astron. Soc. Pac.* **125**, 227 (2013).

20. S. Seager, M. Kuchner, C. A. Hier-Majumder, B. Militzer, Mass-Radius Relationships for Solid Exoplanets, *Astrophys. J.* **669**, 1279 (2007).

21. L. Zeng, D. D. Sasselov, S. B. Jacobsen, Mass-Radius Relation for Rocky Planets Based on PREM, *Astrophys. J.* **819**, 127 (2016).

22. L. Zeng, S. B. Jacobsen, D. D. Sasselov, M. I. Petaev, A. Vanderburg, M. Lopez-Morales, et al., Growth model interpretation of planet size distribution, *Proceedings of the National Academy of Science* **116**, 9723 (2019).

23. A. Santerne, B. Brugger, D.J. Armstrong, V. Adibekyan, J. Lillo-Box, H. Gosselin, et al., An Earth-sized exoplanet with a Mercury-like composition, *Nature Astronomy* **2**, 393 (2018).

24. O. Barragán, D. Gandolfi, F. Dai, J. Livingston, C. M. Persson, T. Hirano, *et al.*, K2-141

b. A 5-M$_\oplus$ super-Earth transiting a K7 V star every 6.7 h, *Astron. Astrophys.* **612**, A95 (2018).

25. L. Malavolta, A. W. Mayo, T. Louden, V. M. Rajpaul, A. S. Bonomo, L. A. Buchhave, et al., An Ultra-short Period Rocky Super-Earth with a Secondary Eclipse and a Neptune-like Companion around K2-141, *Astron. J.* **155**, 107 (2018).

26. D. Jontof-Hutter, J. F. Rowe, J. J. Lissauer, D. C. Fabrycky, E. B. Ford, The mass of the Mars-sized exoplanet Kepler-138 b from transit timing, *Nature* **522**, 321 (2015).

27. J. M. Almenara, R. F. Díaz, C. Dorn, X. Bonfils, S. Udry, Absolute densities in exoplanetary systems: photodynamical modelling of Kepler-138, *Mon. Notices Royal Astron. Soc.* **478**, 460 (2018).

28. M. Gillon, E. Jehin, S. M. Lederer, L. Delrez, J. de Wit, A. Burdanov, V. Van Grootel, et al., Temperate Earth-sized planets transiting a nearby ultracool dwarf star, *Nature* **533**, 221 (2016).

29. M. Gillon, A. H. M. J. Triaud, B.-O. Demory, E. Jehin, E. Agol, K. M. Deck, et al., Seven temperate terrestrial planets around the nearby ultracool dwarf star TRAPPIST-1, *Nature* **542**, 456 (2017).

30. D. E. Smith, M. T. Zuber, R. J. Phillips, S. C. Solomon, S. A. Hauck, F. G. Lemoine, et al., Gravity Field and Internal Structure of Mercury from MESSENGER, *Science*, **336**, 214 (2012).

31. R. A. Marcus, D. Sasselov, L. Hernquist, S. T. Stewart, Minimum Radii of Super-Earths: Constraints from Giant Impacts, *Astrophys. J. Lett.* **712**, L73 (2010).

32. P. Baumeister, S. Padovan, N. Tosi, G. Montavon, N. Nettelmann, J. MacKenzie, et al., Machine-learning Inference of the Interior Structure of Low-mass Exoplanets, *Astrophys. J.* **889**, 42 (2020).

33. J. D. Twickens, et al., B. D. Clarke}, S. T. Bryson, P. Tenenbaum, and H. Wu and J. M. Jenkins, et al., "Photometric analysis in the Kepler Science Operations Center pipeline" in *Software and Cyberinfrastructure for Astronomy*, N. M. Radziwill, A. Bridger, eds. (2010), vol. 7740 of Society of Photo-Optical Instrumentation Engineers (SPIE*) Conference Series*, p. 774023D.

34. J. C. Smith, M. C. Stumpe, J. E. Van Cleve, J. M. Jenkins, T. S. Barclay, M. N. Fanelli, *et al.*, Kepler Presearch Data Conditioning II - A Bayesian Approach to Systematic Error Correction, *Publ. Astron. Soc. Pac.* **124**, 1000 (2012).

35. M. C. Stumpe, J. C. Smith, J. E. Van Cleve, J. D. Twicken, T. S. Barclay, M. N. Fanelli, et al., Kepler Presearch Data Conditioning I —Architecture and Algorithms for Error Correction in Kepler Light Curves, *Publ. Astron. Soc. Pac.* **124**, 985 (2012).

36. M. C. Stumpe, J. C. Smith, J. H. Catanzarite, J. E. Van Cleve, J. M. Jenkins, J. D. Twicken, *et al.*, Multiscale Systematic Error Correction via Wavelet-Based Band-splitting in Kepler Data, *Publ. Astron. Soc. Pac.* **126**, 100 (2014).

37. Mikulski Archive for Space Telescopes, TESS Data Archive, https://archive.stsci.edu/tess (2019).

38. J. M. Jenkins, The Impact of Solar-like Variability on the Detectability of Transiting Terrestrial Planets, *Astrophys. J.* **575**, 493 (2002).

39. J. M. Jenkins, H. Chandrasekaran, S. D. McCauliff, D. A. Caldwell, P. Tenenbaum, J. Li, T. C. Klaus, et al., "Transiting planet search in the Kepler pipeline" in *Software and Cyberinfrastructure for Astronomy*, N. M. Radziwill, A. Bridger, eds. (2010), vol. 7740 of Society of Photo-Optical Instrumentation Engineers (SPIE) Conference Series, p. 77400D.

40. J. D. Twicken, J. H. Catanzarite, B. D. Clarke, F. Girouard, J. M. Jenkins, T. C. Klaus, *et al.*, Kepler Data Validation I—Architecture, Diagnostic Tests, and Data Products for Vetting Transiting Planet Candidates, *Publ. Astron. Soc. Pac.* **130**, 064502 (2018).

41. J. Li, P. Tenenbaum, J. D. Twicken, C. J. Burke, J. M. Jenkins, E. V. Quintana, *et al.*, Kepler Data Validation II-Transit Model Fitting and Multiple-planet Search, *Publ. Astron. Soc. Pac.* **131**, 024506 (2019).

42. A. Savitzky, M. J. E. Golay, Smoothing and differentiation of data by simplified least squares procedures, *Analytical Chemistry* **36**, 1627 (1964).

43. W. H. Press, S. A. Teukolsky, W. T. Vetterling, B. P. Flannery, *Numerical recipes in C++: the art of scientific computing* (2002).

44. J. Korth, "Characterization of extrasolar multi-planet systems by transit timing variation" thesis, Universität zu Köln (2020).

45. S. Grziwa, M. Pätzold, L. Carone, The needle in the haystack: searching for transiting extrasolar planets in CoRoT stellar light curves, *Mon. Notices Royal Astron. Soc.* **420**, 1045 (2012).

46. P. Eigmüller, D. Gandolfi, C. M. Persson, P. Donati, M. Fridlund, S. Csizmadia,*et al.*, K2-

60b and K2-107b. A Sub-Jovian and a Jovian Planet from the K2 Mission, *Astron. J.* **153**, 130 (2017).

47. S. Covino, M. Stefanon, G. Sciuto, A. Fernandez-Soto, G. Tosti, F. M. Zerbi, et al., "REM: a fully robotic telescope for GRB observations" in *Ground-based Instrumentation for Astronomy*, A. F. M. Moorwood, M. Iye, eds. (2004), vol. 5492 of *Society of Photo-Optical Instrumentation Engineers (SPIE) Conference Series*, pp. 1613–1622.

48. D. R. Ciardi, C. A. Beichman, E. P. Horch, S. B. Howell, Understanding the Effects of Stellar Multiplicity on the Derived Planet Radii from Transit Surveys: Implications for Kepler, K2, and TESS, *Astrophys. J.* **805**, 16 (2015).

49. T. M. Brown, N. Baliber, F. B. Bianco, M. Bowman, B. Burleson, P. Conway, *et al.*, Las Cumbres Observatory Global Telescope Network, *Publ. Astron. Soc. Pac* **125**, 1031 (2013).

50. K. A. Collins, J. F. Kielkopf, K. G. Stassun, F. V. Hessman, AstroImageJ: Image Processing and Photometric Extraction for Ultra-precise Astronomical Light Curves, *Astron. J.*, **153**, 77 (2017)

51. K. V. Lester, et al., R. A. Matson, S. B. Howell, E. Furlan, C. L. Gnilka, N. J. Scott, Speckle Observations of TESS Exoplanet Host Stars. II. Stellar Companions at 1-1000 au and Implications for Small Planet Detection, *Astron. J.*, **162**, 75 (2021)

52. R. A. Matson, S. B. Howell, E. P. Horch, M. E. Everett, Stellar Companions of Exoplanet Host Stars in K2, *Astron. J.* **156**, 31 (2018).

53. N. J. Scott, "`Alopeke, Zorro, and NESSI: Three dual-channel speckle imaging

instruments at Gemini-North, Gemini-South, and the WIYN telescopes." In *AAS/Division for Extreme Solar Systems Abstracts* (2019), vol. 51 of AAS/Division for Extreme Solar Systems Abstracts, p. 330.15

54. S. B. Howell, N. J. Scott, R. A. Matson, M. E. Everett, E. Furlan, C. L. Gnilka, et al., The NASA High-Resolution Speckle Interferometric Imaging Program: Validation and Characterization of Exoplanets and Their Stellar Hosts, *Frontiers in Astronomy and Space Sciences*, **8**, 10 (2021)

55. S. B. Howell, M. E. Everett, E. P. Horch, J. G. Winters, L. Hirsch, D. Nusdeo, et al., Speckle Imaging Excludes Low-mass Companions Orbiting the Exoplanet Host Star TRAPPIST-1, *Astrophys. J. Lett.* **829**, L2 (2016).

56. S. B. Howell, M. E. Everett, W. Sherry, E. Horch, D. R. Ciardi, Speckle Camera Observations for the NASA Kepler Mission Follow-up Program, *Astron. J.* **142**, 19 (2011).

57. D. L. Pollacco, A. Collier Cameron, D. J. Christian, C. Hellier, J. Irwin, T. A. Lister, et al., The WASP Project and the SuperWASP Cameras, *Publ. Astron. Soc. Pac.* **118**, 1407 (2006).

58. A. M. S. Smith, WASP Consortium, The SuperWASP exoplanet transit survey, *Contributions of the Astronomical Observatory Skalnate Pleso* **43**, 500 (2014).

59. P. F. L. Maxted, D. R. Anderson, A. Collier Cameron, C. Hellier, D. Queloz, B. Smalley, et al., WASP-41b: A Transiting Hot Jupiter Planet Orbiting a Magnetically Active G8V Star, *Publ. Astron. Soc. Pac.* **123**, 547 (2011).

60. R. A. García, S. Mathur, S. Pires, C. Régulo, B. Bellamy, P. L. Pallé, et al., Impact on

asteroseismic analyses of regular gaps in Kepler data, *Astron. Astrophys.* **568**, A10 (2014).

61. S. Pires, S. Mathur, R. A. García, J. Ballot, D. Stello, K. Sato, Gap interpolation by inpainting methods: Application to ground and space-based asteroseismic data, *Astron. Astrophys.* **574**, A18 (2015).

62. C. Torrence, G. P. Compo, A Practical Guide to Wavelet Analysis., *Bulletin of the American Meteorological Society* **79**, 61 (1998).

63. S. Mathur, R. A. García, C. Régulo, O. L. Creevey, J. Ballot, D. Salabert, et al., Determining global parameters of the oscillations of solar-like stars, *Astron. Astrophys.* **511**, A46 (2010).

64. R. A. García, T. Ceillier, D. Salabert, S. Mathur, J. L. van Saders, M. Pinsonneault, *et al.*, Rotation and magnetism of Kepler pulsating solar-like stars. Towards asteroseismically calibrated age-rotation relations, *Astron. Astrophys.* **572**, A34 (2014).

65. A. McQuillan, T. Mazeh, S. Aigrain, Rotation Periods of 34,030 Kepler Main-sequence Stars: The Full Autocorrelation Sample, *Astrophys. J., Suppl. Ser.* **211**, 24 (2014).

66. T. Ceillier, J. Tayar, S. Mathur, D. Salabert, R. A. García, D. Stello, *et al.*, Surface rotation of Kepler red giant stars, *Astron. Astrophys.* **605**, A111 (2017).

67. A. R. G. Santos, R. A. García, S. Mathur, L. Bugnet, J. L. van Saders, T. S. Metcalfe, *et al.*, Surface Rotation and Photometric Activity for Kepler Targets. I. M and K Main-sequence Stars, *Astrophys. J., Suppl. Ser.* **244**, 21 (2019).

68. B. L. Canto Martins, R. L. Gomes, Y. S. Messias, S. R. de Lira, I. C. Leão, L. A. Almeida, *et al.*, A Search for Rotation Periods in 1000 TESS Objects of Interest, *Astrophys. J.,*

*Suppl. Ser.* **250**, 20 (2020).

69. N. Astudillo-Defru, X. Delfosse, X. Bonfils, T. Forveille, C. Lovis, J. Rameau, Magnetic activity in the HARPS M dwarf sample. The rotation-activity relationship for very low-mass stars through R'(HK), *Astron. Astrophys.* **600**, A13 (2017).

70. M. Mayor, F. Pepe, D. Queloz, F. Bouchy, G. Rupprecht, G. Lo Curto, *et al.*, Setting New Standards with HARPS, *The Messenger* **114**, 20 (2003).

71. X. Bonfils, X. Delfosse, S. Udry, T. Forveille, M. Mayor, C. Perrier, *et al.*, The HARPS search for southern extra-solar planets. XXXI. The M- dwarf sample, *Astron. Astrophys.* **549**, A109 (2013).

72. C. Lovis, F. Pepe, A new list of thorium and argon spectral lines in the visible, *Astron. Astrophys.* **468**, 1115 (2007).

73. G. Lo Curto, F. Pepe, G. Avila, H. Boffin, S. Bovay, B. Chazelas, *et al.*, HARPS Gets New Fibres After 12 Years of Operations, *The Messenger* **162**, 9 (2015).

74. N. Astudillo-Defru, T. Forveille, X. Bonfils, D. Ségransan, F. Bouchy, X. Delfosse, et al., The HARPS search for southern extra-solar planets. XLI. A dozen planets around the M dwarfs GJ 3138, GJ 3323, GJ 273, GJ 628, and GJ 3293, *Astron. Astrophys.* **602**, A88 (2017).

75. M. Zechmeister, A. Reiners, P. J. Amado, M. Azzaro, F. F. Bauer, V. J. S. Béjar, et al., Spectrum radial velocity analyser (SERVAL). High-precision ra- dial velocities and two alternative spectral indicators, *Astron. Astrophys.* **609**, A12 (2018).

76. S. W. Yee, E. A. Petigura, K. von Braun, Precision Stellar Characterization of FGKM Stars

using an Empirical Spectral Library, *Astrophys. J.* **836**, 77 (2017).

77. T. Hirano, F. Dai, D. Gandolfi, A. Fukui, J. H. Livingston, K. Miyakawa, et al., Exoplanets around Low-mass Stars Unveiled by K2, *Astron. J.* **155**, 127 (2018).

78. A. W. Mann, T. Dupuy, A. L. Kraus, E. Gaidos, M. Ansdell, M. Ireland, A. C. Rizzuto, et al., How to Constrain Your M Dwarf. II. The Mass-Luminosity-Metallicity Relation from 0.075 to 0.70 Solar Masses, *Astrophys. J.* **871**, 63 (2019).

79. K. G. Stassun, G. Torres, Evidence for a Systematic Offset of -80 $\mu$as in the Gaia DR2 Parallaxes, *Astrophys. J.* **862**, 61 (2018).

80. A. Antoniadis-Karnavas, S. G. Sousa, E. Delgado-Mena, N. C. Santos, G. D. C. Teixeira, V. Neves, ODUSSEAS: a machine learning tool to derive effective temperature and metallicity for M dwarf stars, *Astron. Astrophys.* **636**, A9 (2020).

81. T. D. Morton, isochrones: Stellar model grid package, Astrophysics Source Code Library, ascl:1503.010 (2015).

82. J. Choi, A. Dotter, C. Conroy, M. Cantiello, B. Paxton, B. D. Johnson, Mesa Isochrones and Stellar Tracks (MIST). I. Solar-scaled Models, *Astrophys. J.* **823**, 102 (2016).

83. Gaia Collaboration, T. Prusti, J. H. J. de Bruijne, A. G. A. Brown, A. Vallenari, C. Babusiaux, et al., The Gaia mission, *Astron. Astrophys.* **595**, A1 (2016).

84. M. F. Skrutskie, R. M. Cutri, R. Stiening, M. D. Weinberg, S. Schneider, J. M. Carpenter, et al., The Two Micron All Sky Survey (2MASS), *Astron. J.* **131**, 1163 (2006).

85. F. Feroz, M. P. Hobson, E. Cameron, A. N. Pettitt, Importance Nested Sampling and the

MultiNest Algorithm, *ArXiv e-prints* arXiv:1306.2144 (2013).

86. J. C. Morales, J. Gallardo, I. Ribas, C. Jordi, I. Baraffe, G. Chabrier, The Effect of Magnetic Activity on Low-Mass Stars in Eclipsing Binaries, *Astrophys. J.* **718**, 502 (2010).

87. A. L. Kraus, R. A. Tucker, M. I. Thompson, E. R. Craine, L. A. Hillenbrand, The Mass-Radius(-Rotation?) Relation for Low-mass Stars, *Astrophys. J.* **728**, 48 (2011).

88. G. Torres, Fundamental properties of lower main-sequence stars, *Astronomische Nachrichten* **334**, 4 (2013).

89. A. W. Mann, G. A. Feiden, E. Gaidos, T. Boyajian, K. von Braun, How to Constrain Your M Dwarf: Measuring Effective Temperature, Bolometric Luminosity, Mass, and Radius, *Astrophys. J.* **804**, 64 (2015).

90. B. Stelzer, A. Marino, G. Micela, J. López-Santiago, C. Liefke, The UV and X-ray activity of the M dwarfs within 10 pc of the Sun, *Mon. Notices Royal Astron. Soc.* **431**, 2063 (2013).

91. B. S. Gaudi, J. N. Winn, Prospects for the Characterization and Confirmation of Transiting Exoplanets via the Rossiter-McLaughlin Effect, *Astrophys. J.* **655**, 550 (2007).

92. T. Bensby, S. Feltzing, I. Lundström, Elemental abundance trends in the Galactic thin and thick disks as traced by nearby F and G dwarf stars, *Astron. Astrophys.* **410**, 527 (2003).

93. S. A. Barnes, A Simple Nonlinear Model for the Rotation of Main-sequence Cool Stars. I. Introduction, Implications for Gyrochronology, and Color-Period Diagrams, *Astrophys. J.* **722**, 222 (2010).

94. S. A. Barnes, Y.-C. Kim, Angular Momentum Loss from Cool Stars: An Empirical Expression and Connection to Stellar Activity, *Astrophys. J.* **721**, 675 (2010).

95. R. Angus, T. D. Morton, D. Foreman-Mackey, J. van Saders, J. Curtis, S. R. Kane, *et al.*, Toward Precise Stellar Ages: Combining Isochrone Fitting with Empirical Gyrochronology, *Astron. J.* **158**, 173 (2019).

96. B. R. Jørgensen, L. Lindegren, Determination of stellar ages from isochrones: Bayesian estimation versus isochrone fitting, *Astron. Astrophys.* **436**, 127 (2005).

97. M. Zechmeister, M. Kürster, The generalised Lomb-Scargle periodogram. A new formalism for the floating-mean and Keplerian periodograms, *Astron. Astrophys.* **496**, 577 (2009).

98. K. A. Murdoch, J. B. Hearnshaw, M. Clark, A Search for Substellar Companions to Southern Solar-Type Stars, *Astrophys. J.* **413**, 349 (1993).

99. A. P. Hatzes, R. Dvorak, G. Wuchterl, P. Guterman, M. Hartmann, M. Fridlund, et al., An investigation into the radial velocity variations of CoRoT-7, *Astron. Astrophys.* **520**, A93 (2010).

100. A. P. Hatzes, M. Fridlund, G. Nachmani, T. Mazeh, D. Valencia, G. Hébrard, *et al.*, The Mass of CoRoT-7b, *Astrophys. J.* **743**, 75 (2011).

101. A. P. Hatzes, *The Doppler Method for the Detection of Exoplanets* (IOP Publishing, Bristol, UK 2019), doi:10.1088/2514-3433/ab46a3.

102. J. A. Carter, J. N. Winn, Parameter Estimation from Time-series Data with Correlated Errors: A Wavelet-based Method and its Application to Transit Light Curves, *Astrophys. J.*

**704**, 51 (2009).

103. S. Csizmadia, et al., The power of wavelets in analysis of transit and phase curves in presence of stellar variability and instrumental noise I. Method and validation, *arXiv e-prints*, arXiv:2108.11822 (2021)

104. A. P. Hatzes, The detection of Earth-mass planets around active stars. The mass of Kepler-78b, *Astron. Astrophys.* **568**, A84 (2014).

105. A. Claret, Limb and gravity-darkening coefficients for the TESS satellite at several metallicities, surface gravities, and microturbulent velocities, *Astron. Astrophys.* **600**, A30 (2017).

106. B. Croll, Markov Chain Monte Carlo Methods Applied to Photometric Spot Modeling, *Publ. Astron. Soc. Pac.* **118**, 1351 (2006).

107. E. B. Ford, Improving the Efficiency of Markov Chain Monte Carlo for Analyzing the Orbits of Extrasolar Planets, *Astrophys. J.* **642**, 505 (2006).

108. S. Csizmadia, A. Hatzes, D. Gandolfi, M. Deleuil, F. Bouchy, M. Fridlund, et al., Transiting exoplanets from the CoRoT space mission. XXVIII. CoRoT-33b, an object in the brown dwarf desert with 2:3 commensurability with its host star, *Astron. Astrophys.* **584**, A13 (2015).

109. J. N. Winn, Transits and Occultations, *arXiv e-prints* arXiv:1001.2010 (2010).

110. F. Dai, J. N. Winn, D. Gandolfi, S. X. Wang, J.~K. Teske, J. Burt, et al., The Discovery and Mass Measurement of a New Ultra-short-period Planet: K2-131b, *Astron. J.* **154**, 226 (2017).

111. F. Dai, K. Masuda, J. N. Winn, L. Zeng, Homogeneous Analysis of Hot Earths: Masses, Sizes, and Compositions, *Astrophys. J. Lett.* **883**, 79 (2019).

112. R. Caruana, Multitask Learning, *Machine Learning* **28**, 41 (1997).

113. C. M. Bishop, *Neural Networks for Pattern Recognition* (Oxford University Press Inc., New York, 1995).

114. C. Dorn, J. Venturini, A. Khan, K. Heng, Y. Alibert, R. Helled, et al., A generalized Bayesian inference method for constraining the interiors of super Earths and sub-Neptunes, *Astron. Astrophys.* **597**, A37 (2017).

115. C. Sotin, O. Grasset, A. Mocquet, Mass radius curve for extrasolar Earth-like planets and ocean planets, *Icarus* **191**, 337 (2007).

116. E. E. Salpeter, H. S. Zapolsky, Theoretical High-Pressure Equations of State including Correlation Energy, *Physical Review* **158**, 876 (1967).

117. W. Kley, R. P. Nelson, Planet-Disk Interaction and Orbital Evolution, *Astron. Astrophys.* **50**, 211 (2012).

118. F. A. Rasio, E. B. Ford, Dynamical instabilities and the formation of extrasolar planetary systems, *Science* **274**, 954 (1996).

119. A. M. Dziewonski, D. L. Anderson, Preliminary reference Earth model, *Physics of the Earth and Planetary Interiors* **25**, 297 (1981).

120. P. Goldreich, S. Soter, Q in the Solar System, *Icarus* **5**, 375 (1966).

121. K. C. Patra, J. N. Winn, M. J. Holman, L. Yu, D. Deming, F. Dai, The Apparently Decaying

Orbit of WASP-12b, *Astron. J.* **154**, 4 (2017).

122. Planetary Systems Table, *NASA Exoplanet Archive*, DOI: 10.26133/NEA12

123. E. M. R. Kempton, J. L. Bean, D. R. Louie, D. Deming, D. D. B. Koll, M. Mansfield, et al., A Framework for Prioritizing the TESS Planetary Candidates Most Amenable to Atmospheric Characterization, *Publ. Astron. Soc. Pac.* **130**, 114401 (2018).

124. J. Chen, D. Kipping, Probabilistic Forecasting of the Masses and Radii of Other Worlds, *Astrophys. J.* **834**, 17 (2017).

125. P. W. Sullivan, J. N. Winn, Z. K. Berta-Thompson, D. Charbonneau, D. Deming, C. D. Dressing, et al., The Transiting Exoplanet Survey Satellite: Simulations of Planet Detections and Astrophysical False Positives, *Astrophys. J.* **809**, 77 (2015).

126. K. G. Stassun, R. J. Oelkers, M. Paegert, G. Torres, J. Pepper, N. De Lee, *et al.*, The Revised TESS Input Catalog and Candidate Target List, *Astron. J.* **158**, 138 (2019).

127. R. M. Cutri, M. F. Skrutskie, S. van Dyk, C. A. Beichman, J. M. Carpenter, T. Chester, et al., VizieR Online Data Catalog: 2MASS All-Sky Catalog of Point Sources (Cutri+ 2003), *VizieR Online Data Catalog* p. II/246 (2003).

**Acknowledgments:**

We acknowledge use of observations from the LCOGT network. We made use of data from the European Space Agency (ESA) mission Gaia (https:// www.cosmos.esa.int/gaia), processed by the Gaia Data Processing and Analysis Consortium (DPAC, https://www.cosmos.esa.int/web/gaia/dpac/consortium). Funding for the DPAC has been provided by national institutions, in particular the institutions participating in the Gaia Multilateral Agreement. We acknowledge the use of public TESS Alert data from pipelines at the TESS Science Office and at the TESS Science Processing Operations Center. Supported by the KESPRINT collaboration, an international consortium devoted to the characterization and research of exoplanets discovered with space-based missions. Some of the observations were made at Gemini South using the High-Resolution Imaging instrument Zorro, funded by the NASA Exoplanet Exploration Program and built at the NASA Ames Research Center by Steve B. Howell, Nic Scott, Elliott P. Horch, and Emmett Quigley. Zorro was mounted on the Gemini South telescope of the international Gemini Observatory, a pro-gram of NSFs OIR Lab, which is managed by the Association of Universities for Research in Astronomy (AURA) under a cooperative agreement with the National Science Foundation. on behalf of the Gemini partnership: the National Science Foundation (United States), National Research Council (Canada), Agencia Nacional de Investigacin y Desarrollo (Chile), Ministe-rio de Ciencia, Tecnologa e Innovacin (Argentina), Ministrio da Ciłncia, Tecnologia, Inovaes e Comunicaes (Brazil), and Korea Astronomy and Space Science Institute (Republic of Korea). This is University of Texas Center for Planetary Systems Habitability Contribution #0041. DG and LMS gratefully acknowledge financial support from the Cassa di Risparmio di Torino foundation under Grant No. 2018.2323 "Gaseous or rocky? Unveiling the nature of small worlds.

**Funding:**

K.W.F.L., Sz.C., M.E., S.G., A.P.H. and H.R were supported by Deutsche Forschungsgemeinschaft grants PA525/18-1, PA525/19-1, PA525/20-1, HA3279/12-1 and RA714/14-1 within the DFG Schwerpunkt SPP 1992, Exploring the Diversity of Extrasolar

Planets. Sz.C. is supported by Deutsche Forschungsgemeinschaft Research Unit 2440: 'Matter Under Planetary Interior Conditions: High Pressure Planetary and Plasma Physics'. T.H. was supported by JSPS KAKENHI Grant Numbers JP19K14783, N.N. was supported by JSPS KAKENHI Grant Numbers JP18H01265 and JP18H05439, and JST PRESTO Grant Number JPMJPR1775. J.L. is supported by JSPS KAKENHI Grant Number JP20K14518. R.A.G. acknowledges support from PLATO and GOLF CNES grants. P.K. acknowledges support from the MSMT grant LTT20015. S.M. acknowledges support by the Spanish Ministry of Science and Innovation with the Ramon y Cajal fellowship number RYC-2015-17697 and grant number PID2019-107187GB-I00. N.S. was supported by Fundaçao para a Ciência e a Tecnologia through national funds and by FEDER through COMPETE2020 - Programa Operacional Competitivida e Internacionalização grants UID/FIS/04434/2019; UIDB/04434/2020; UIDP/04434/2020; PTDC/FIS-AST/32113/2017 & POCI-01-0145-FEDER-032113; PTDC/FIS-AST/28953/2017 & POCI-01-0145-FEDER-028953. X.D. and G.G acknowledge funding in the framework of the Investissements d'Avenir program (ANR-15-IDEX-02) - Origin of Life project of the Univ. Grenoble-Alpes. J.R.M acknowledges grants from CNPq, CAPES and FAPERN Brazilian agencies. N. A.-D. acknowledges support from FONDECYT project 3180063. Resources for the production of the SPOC data products were provided by the NASA High-End Computing (HEC) Program through the NASA Advanced Supercomputing (NAS) Division at Ames Research Center. S. A. is supported by the Danish National Research Foundation (Grant agreement no.: DNRF106)

**Author contributions:**

KWFL contributed to the planet detection, transit light curve analysis, gyrochonology age estimate, tidal evolution calculations, and led the writing of the paper. SzC performed the joint light curve and RV analysis using TLCM. NAD and XB led the HARPS RV follow-up program, reduced and analyzed the RV data. DG performed the frequency analysis of the RV and activity indicators. FD, OB, APH and RL analyzed the RV data. SP determined the interior composition of the planet and modelled the precursor gaseous planet. ME performed the REM observations and analysis. ME and AMSS derived the light curve dilution factor. CH, KWFL, SM and RAG determined the stellar rotation period using WASP data. SBH analyzed the speckle imaging data. TH and MF performed the spectral analysis using SpecMatch-Emp and TH derived the stellar parameters using an MCMC simulation. JL performed stellar characterization using isochrone fitting. FM performed the LCOGT and HARPS observations. KAC coordinated the TESS SG1 working group. KAC and PL analyzed the ground-based photometric observations. NCS performed spectral characterization using ODUSSEAS. MCJ determined the Galactic space velocities of the star. KWFL and SR calculated the emission spectroscopy metric. JK performed TTV analysis. JC, PhE and SG analyzed the light curve for planet detection. EGu analyzed the activity indicators. GRR, RV, DWL, SS, JNW and JMJ led and organized the TESS mission, including the observations, processing of the data, working groups coordination, target selection, and dissemination of the data products. EHM, MV and JES are members of the TESS POC who coordinated and scheduled the TESS science observations and conducted instrument planning. RM, JCS, JDT are members of the SPOC who performed data calibration, light curve production and transit planet detection. DC, JC, SNQ are members of the TSO who reviewed the data and performed planet candidate vetting. JMA, EA, FB, DC,

JRDM, XD, RFD, RD, PF, TF, GG, CL, CM, FP, DS, SU are Co-Is of the proposal which provided the HARPS observations of the target and supported the interpretation of results. SA, PC, AC, WDC, EGo, IG, PK, EK, JJL, NN, HLMO, EP, CMP, HR, LS, JS, VVE are part of the KESPRINT consortium and contributed to the interpretation of the results. All authors contributed to the preparation of the paper.

**Competing interests:**

We declare no competing interests.

**Data and materials availability:**

The TESS photometric observations are available at the Mikulski Archive for Space Telescopes (MAST) at https://exo.mast.stsci.edu under target name TOI 731.01. The raw HARPS spectroscopic data are available on the ESO Science Archive Facility http://archive.eso.org/cms.html under ESO program IDs 072.C-0488, 082.C-0718, 183.C-0437 (PIs: M. Mayor and X. Bonfils) and 1102.C-0339 (PI: X. Bonfils). The ground-based photometry obtained by the LCO telescope and REM, and the Gemini imaging data are available on the Exoplanet Follow-up Observing Program (ExoFOP) website https://exofop.ipac.caltech.edu/tess/ under target name TOI 731.01. The raw Gemini data are available at https://archive.gemini.edu/searchform under Program ID GS-2021A-LP-105. The archival WASP data are available on the NASA Exoplanet Archive https://exoplanetarchive.ipac.caltech.edu/docs/SuperWASPMission.html under object name GJ 367. Our reduced RVs and activity indices are listed in Tables S1-S2 and in machine-readable form in Data S1-S2. The TRANSIT AND LIGHT CURVE MODELLER (TLCM) is available at

http://www.transits.hu/ .

**Supplementary Materials**

Materials and Methods

Supplementary Text

Figs. S1 to S10

Tables S1 to S3

Data S1 to S2

References (*32-127*)

**Table 1: Properties of host star GJ 367 and planet GJ 367b.** The stellar parameters were derived from the spectral analysis of the HARPS data (*9*). Planet parameters were obtained from the joint model fitting of the TESS photometry and HARPS RVs (*9*). Reported values are the medians of the posterior probability distributions with uncertainties of the 34th and 68th percentiles of those distributions.

| Parameter | Value |
|---|---|
| **Star GJ 367 (TOI 731)** | |
| Right ascension (J2000 equinox) | 09h44m29s.84 |
| Declination (J2000 equinox) | -45°46 ′35″.43 |
| TESS-band magnitude | 8.032 ± 0.007 |
| *V*-band magnitude | 10.153 ± 0.044 |
| Parallax[*] (milliarcsecond, mas) | 106.272 ± 0.056 |
| Distance, $d$ (pc) | 9.410 ± 0.005 |
| Effective Temperature, $T_{\text{eff}}$ (K) | 3522 ± 70 |
| Stellar mass, $M_s$ ($M_\odot$) | 0.454 ± 0.011 |
| Stellar radius, $R_s$ ($R_\odot$) | 0.457 ± 0.013 |
| Stellar density, $\rho_s$ ($\rho_\odot$) | $4.76^{+0.43}_{-0.39}$ |
| Metallicity, [Fe/H] | -0.01 ± 0.12 |
| Surface gravity, log $g$ | 4.777 ± 0.026 |
| Luminosity, $L_s$ ($L_\odot$) | 0.0288 ± 0.0027 |
| Spectral type | M1.0V |
| **Planet GJ 367b** | |
| Epoch, $T_0$ [barycentric Julian date, BJD$_{\text{TDB}}$] | 2458544.1348 ± 0.0004 |
| Orbital Period, $P$ (days) | $0.321962^{+0.000010}_{-0.000012}$ |
| Planet-to-star radius ratio, $R_p/R_s$ | $0.0143^{+0.0096}_{-0.0010}$ |
| Scaled orbital semi-major axis, $a/R_s$ | $3.41^{+0.06}_{-0.07}$ |
| Impact parameter, $b$ | $0.55^{+0.03}_{-0.04}$ |
| Radial velocity semi-amplitude[†], $K$ (cm s⁻¹) | 79.8 ± 11.0 |
| Systemic radial velocity[‡], $v_\gamma$ (km s ⁻¹) | 47.9258 ± 0.0003 |
| Eccentricity, $e$ | 0 |
| Transit duration, $T_{14}$ (min) | $36.9^{+1.0}_{-0.9}$ |
| Orbital semi-major axis, $a$ (au) | 0.0071 ± 0.0002 |
| Orbital inclination, $i$ (º) | 80.75 ± 0.64 |
| Planet mass, $M_p$ ($M_\oplus$) | 0.546 ± 0.078 |
| Planet radius, $R_p$ ($R_\oplus$) | 0.718 ± 0.054 |
| Planet bulk density, $\rho_p$ (g cm$^{-3}$) | $8.106 \pm 2.165$ |
| Equilibrium Dayside Temperature[§], $T_{eq}$ (K) | |
| —Earth-like bond albedo ($A_b = 0.3$) | $1597 \pm 39$ |
| —Zero bond albedo | $1745 \pm 43$ |

[*]A correction of +61 mas was applied to the Gaia parallax (*9*)

[†]Radial velocity induced by the orbiting planet

[‡]Radial velocity of the star-planet system with respect to the observer. The uncertainty only reflects the internal precision of HARPS and does not account for systematic effects such as gravitational redshift.

[§]Assuming no atmospheric circulation

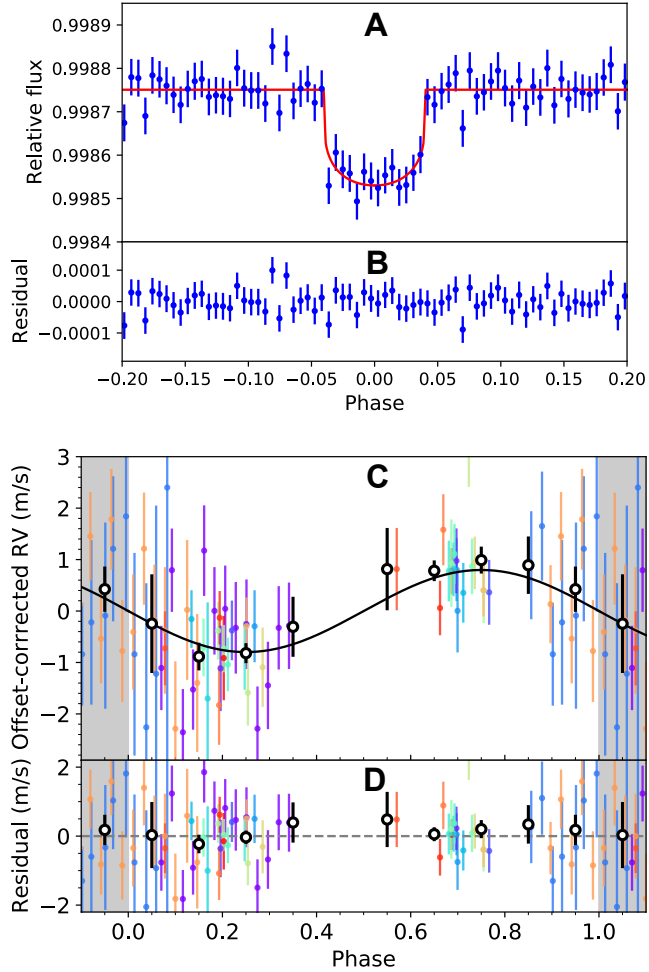

Figure 1: **Phase-folded radial velocity and light curve of GJ 367. (A)** Phase-folded, 2.6-minute binned TESS light curve (blue circles) of GJ 367 with the best-fitting transit model (red line). **(B)** The residuals of the light curve. A noise correction model has been applied to the data (*9*). **(C)** Phase-folded HARPS radial velocity data for GJ 367. Different color dots correspond to different corrections applied to the RV model (*9*). Black open circles are the RV data phase-binned in intervals of 0.10. The solid black line shows the best-fitting RV model, which has a semi-amplitude of $79.8 \pm 11.0$ cm s$^{-1}$. **(D)** The corresponding residuals of the RV data. In panels C and D, the RV orbital phase limits extend beyond phase 0–1 (shaded grey regions) so the first and last data points are duplicated.

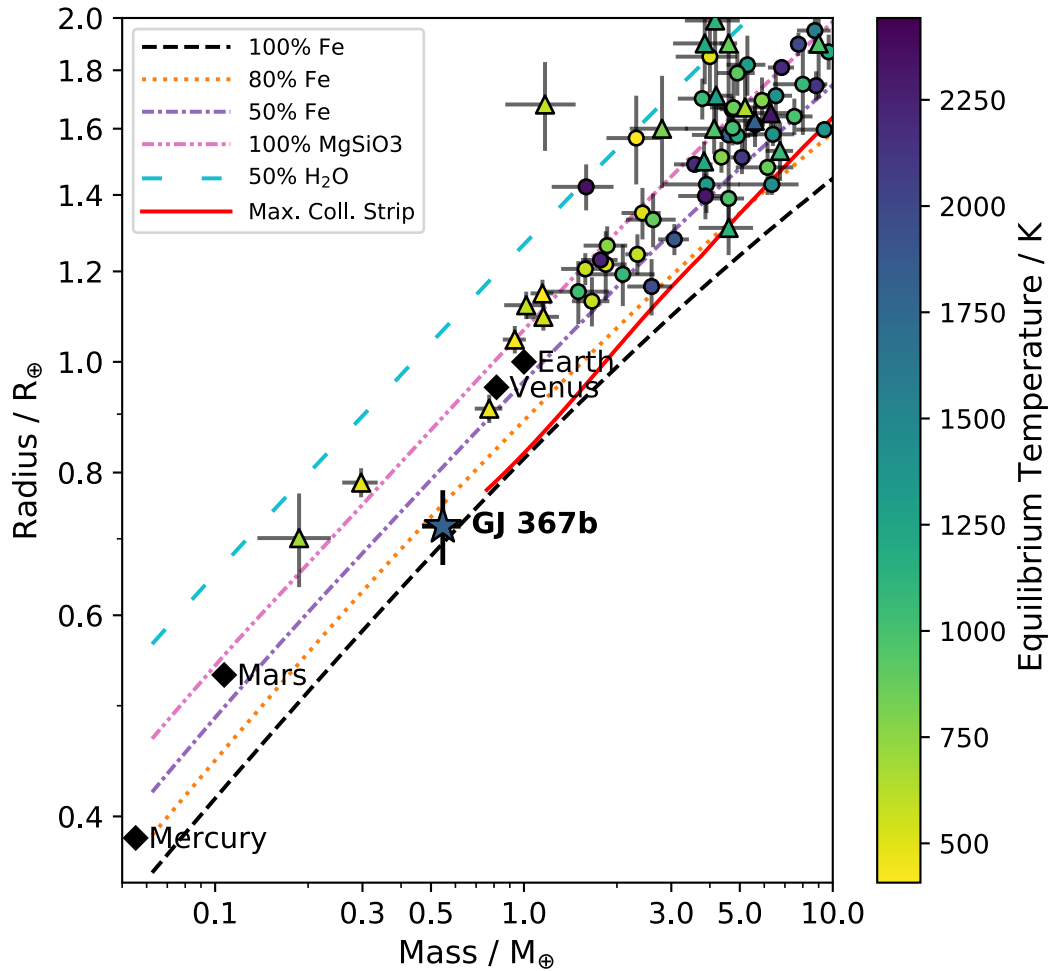

**Figure 2: Masses and radii of small planets (<2 $R_\oplus$) that have both quantities measured with precision ≤ 30%** (*9*). Symbols indicate masses determined with RVs (circles) and transit timing variations (triangles), Solar System planets (diamonds) and GJ 367b (star). Error bars show 1-sigma uncertainties. Exoplanet symbols are color-coded according to the equilibrium dayside temperatures (color bar). Theoretical mass-radius relations for two-layer rocky planets (*22*) are shown with lines corresponding to different core mass fractions. These cores consist of pure iron, pure rock (100% MgSiO3), or a two-layer core with a mixture of iron and rock or rock and $H_2O$ ice, as indicated in the legend. The red solid line denotes the lower limit on planet radius after collisional stripping (*31*). GJ 367b is likely an iron-dominated planet.

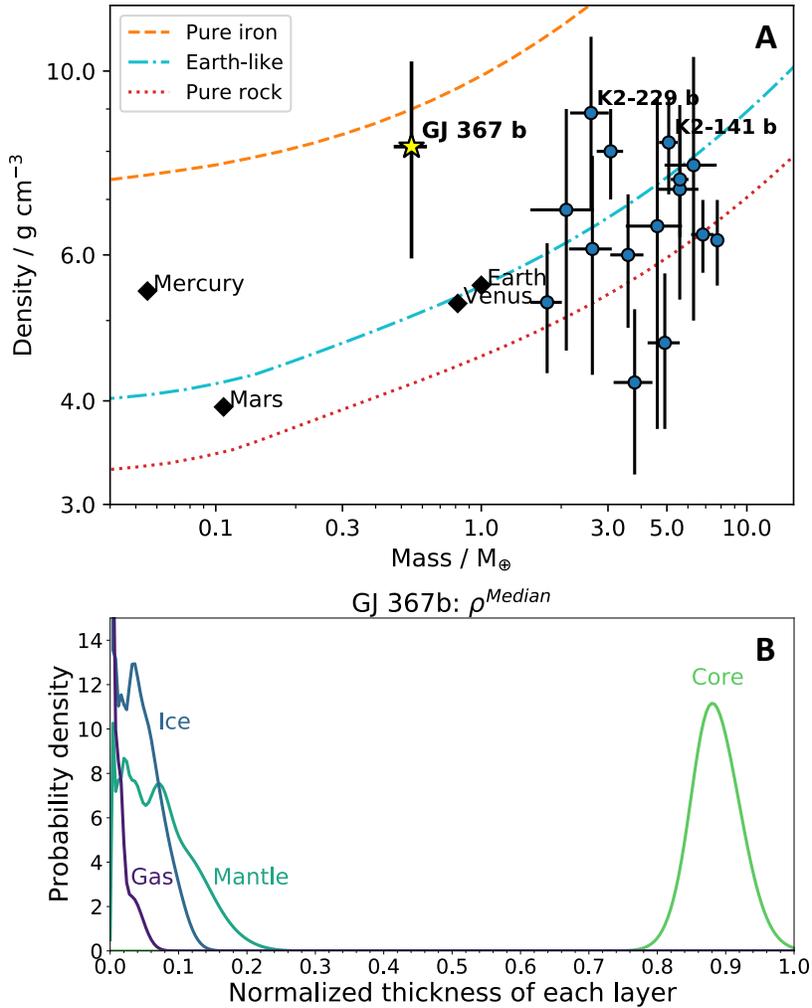

**Figure 3: Bulk composition of USP planets and predicted interior structure of GJ 367b.** **(A)** Mass-density diagram for USP ($P_{orb}$ < 1 day) exoplanets with low mass (<10 $M_\oplus$) and measured mass precisions ≤ 30%. Inner Solar System planets are shown as black diamonds. Planet interior composition models (*22*) are shown with lines indicated in the legend. The bulk densities of low mass USP planets are usually consistent with terrestrial compositions (pure rock or Earth-like). GJ 367b is more consistent with pure iron and an interior similar to Mercury. **(B)** The predicted relative thicknesses of each interior layer of GJ 367b from the neural network model (*32*). The core is assumed to be a liquid a Fe-FeS alloy. The mantle is assumed to be composed of olivine

and orthopyroxene enstatite in the upper mantle and bridgmanite and magnesiowüstite in the lower mantle. The ice layer is assumed to be water ice VII and the gas layer consists of hydrogen and helium. The interior composition of GJ 367b was computed using the median mass and radius measurements (corresponding to the derived median planet density $\rho^{\text{Median}} = 8.106$ g cm$^{-1}$). We infer an iron core filling $86 \pm 5\%$ of the planet radius, with less than 1% ice and gas, similar to the interior of Mercury which has an iron core radius fraction of $83 \pm 2\%$ (*30*). If we take the lowest density of GJ 367b permitted by the observations, 5.941 g cm$^{-1}$, the planet iron core radius fraction is still higher than Earth's (Figure S7).

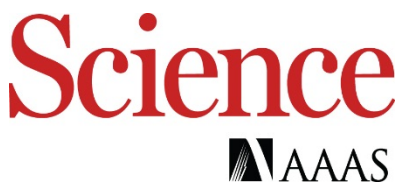

Supplementary Materials for

**GJ 367b: A dense ultra-short period sub-Earth planet transiting a nearby red dwarf star**

Kristine W. F. Lam*, Szilárd Csizmadia, Nicola Astudillo-Defru, Xavier Bonfils, Davide Gandolfi, Sebastiano Padovan, Massimiliano Esposito, Coel Hellier, Teruyuki Hirano, John Livingston, Felipe Murgas, Alexis M. S. Smith, Karen A. Collins, Savita Mathur, Rafael A. Garcia, Steve B. Howell, Nuno C. Santos, Fei Dai, George R. Ricker, Roland Vanderspek, David W. Latham, Sara Seager, Joshua N. Winn, Jon M. Jenkins, Simon Albrecht, Jose M.Almenara, Etienne Artigau, Oscar Barragán, François Bouchy, Juan Cabrera, David Charbonneau, Priyanka Chaturvedi, Alexander Chaushev, Jessie L. Christiansen, William D. Cochran, José-Renan De Meideiros, Xavier Delfosse, Rodrigo F. Díaz ,René Doyon, Philipp Eigmüller, Pedro Figueira, Thierry Forveille, Malcolm Fridlund, Guillaume Gaisné, Elisa Goffo, Iskra Georgieva, Sascha Grziwa, Eike Guenther, Artie P. Hatzes, Marshall C. Johnson, Petr Kabáth, Emil Knudstrup, Judith Korth, Pablo Lewin, Jack J. Lissauer, Christophe Lovis, Rafael Luque, Claudio Melo, Edward H. Morgan, Robert Morris, Michel Mayor, Norio Narita, Hannah L. M. Osborne, Enric Palle, Francesco Pepe, Carina M. Persson, Samuel N. Quinn, Heike Rauer, Seth Redfield, Joshua E. Schlieder, Damien Ségransan, Luisa M. Serrano, Jeffrey C. Smith, Ján Šubjak, Joseph D. Twicken, Stéphane Udry, Vincent Van Eylen, Michael Vezie

*Correspondence to: Kristine.Lam@dlr.de

This PDF file includes:

**Materials and Methods**

<u>TESS observations and transit search</u>

GJ 367 (TIC 34068865 or TOI-731.01) was observed by TESS (*10*) during Sector 9 (28 February 2019 – 25 March 2019) on CCD 1 of Camera 3 at a cadence of 2 minutes. The 2-minute cadence data were processed with the Science Processing Operations Center (SPOC) (*11*) pipeline to produce Simple Aperture Photometry (SAP) (*33*) and Presearch Data Conditioning (PDCSAP) light curves, where the latter are corrected for known instrumental systematics (*34–36*). The TESS data was downloaded from Mikulski Archive for Space Telescopes (MAST) (*37*).

The SPOC Transiting Planet Search (TPS) (*38, 39*) and Data Validation (DV) (*40, 41*) pipelines searched for planets in the resulting light curves. The pipeline detected a transit signal with a period of 0.32 day and a transit depth of 222 ppm. The DST (Détection Spécialisée de Transits) (*12*) algorithm was also used for transit searches. The algorithm first removes the variability in the PDCSAP light curve using a Savitzky-Golay filter (*42, 43*). It then uses a parabolic transit function to perform transit searches. A transit signal was detected with a period of P = 0.321966 ± 0.000031 day, a mid-transit time of Barycentric Julian Date (BJD) = 2458544.1347 ± 0.0013, a transit depth of 293.74 ± 14.41 ppm, and a transit duration of 0.48 ± 0.01 hour. We iteratively ran the algorithm to search for additional periodic transit signals in the light curve but found no further significant detection beyond the 0.32 day signal. We used PYTTV (Python Tool for Transit Variations) (*44*) to extract the transit times of individual transits and found no evidence for transit timing variations. The detection was independently confirmed with the EXOTRANS pipeline (*45*) with a signal detection efficiency (SDE) value of 22, higher than the SDE limit of 9.

There are 25 sources within 30 arcsec of the target star listed in the Gaia Data Release 2 (DR2) catalogue. We therefore performed tests to check for possible astrophysical false positives. As a first test, we downloaded the target pixel files from MAST and used an existing reduction pipeline (*46*) to apply varying aperture sizes  to extract the TESS light curve. We compared the transit depths of each light curve as a function of the aperture sizes. If the transit signal originated from a neighbouring contaminant instead of GJ 367, we would expect the transit depth to increase with the aperture size used for photometry reduction. Figure S1 shows the phase-folded transit light curves of GJ 367b extracted using 3 × 3 pixels (Figure S1A), 4 × 4 pixels (Figure S1B) and 9 × 9 pixels (Figure S1C) apertures. The transit signal of GJ 367b is detected in all light curves and there is no correlation between the transit depths and aperture sizes between light curves. This indicates that the transit signal is likely due to a transiting planet.

<u>Ground-based photometric observations</u>

Photometric observations of GJ 367 were carried out on three consecutive nights (8, 9 and 10 June 2019) with the Rapid Eye Mount (REM) robotic telescope (*47*), located at the ESO La Silla observatory. Each night the target was monitored for ~2.5 hours, bracketing a transit of GJ 367b. We used the REM Optical Slitless Spectrograph 2 (ROSS2), an optical simultaneous multi-channel imaging camera, which provides images in four different bands (Sloan Digital Sky Survey g', r', i', z') onto four quadrants of the same 2k × 2k CCD detector. The images have a field of view of 9.1 × 9.1 arcmin, and a plate scale of 0.58 arcsec pixel$^{-1}$. To avoid saturation of our target in the i' band, we fixed the exposure time to 10 s. All images were bias and flat-field corrected via standard procedures.

By stacking all the images, separately for each night and band, we obtained higher Signal-to-Noise ratio (S/N) images that we used to estimate the contamination by nearby stars in the TESS aperture, as it affects the estimation of the transit depth and planet radius. We derive the contamination factor by comparing the total flux within the TESS aperture before and after masking all the nearby stars. We convolved the REM images with the TESS PSF, assumed to be a 2D Gaussian with width = 22.4 arcsec. We also allowed for uncertainties in the position (0.25 TESS pixels) and orientation (10 degrees) of the TESS aperture on our REM images. The contamination factor in the TESS band was estimated to be $9.5 \pm 1.2\%$, as a weighted average of the values obtained in the r', i' and z' bands. This is consistent with the crowding metric (CROWDSAP) estimated by the SPOC pipeline. The CROWDSAP metric measures the ratio of target flux to total flux in the optimal apertures. The CROWDSAP of GJ 367 is 0.903, equivalent to 9.7% contamination in the TESS light curve. The planetary transit depth appears shallower if the light curve is contaminated by flux from nearby stars *(48)*. For GJ 367, the inferred planet radius was therefore underestimated by about 5% in the pipeline results. The uncertainty in the contamination factor propagates to a radius uncertainty of less than 1%, negligible in determining the planet radius. The contamination factor is used in our transit model to derive the true planet radius (Table 1).

In order to investigate the possibility of a false positive detection, we further observed a full transit of GJ 367b in Panoramic Survey Telescope and Rapid Response System (Pan-STARRS) z-short band on Universal Time (UT) 2020 May 12 for 184 minutes from the Las Cumbres Observatory Global Telescope (LCOGT) *(49)* 1-m network node at Siding Spring Observatory. The LCOGT SINISTRO cameras have an image scale of 0.389" per pixel, giving a 26' × 26' field of view. Light curves of the target and nearby stars were extracted using ASTROIMAGEJ *(50)*. The transit depth of GJ 367b obtained by the SPOC pipeline is 242ppm, which is generally too shallow to reliably detect from the ground. However, the SPOC photometric aperture generally extends ~1' from the target star, so we checked for possible nearby eclipsing binaries (NEBs) out to 2.5 arcminutes from the target star to account for possible contamination from the wings of neighboring star PSFs. A nearby star that is fully blended with the TESS photometric aperture that is up to 9.1 magnitudes fainter in the TESS-band could produce the TESS detection. Our search ruled out NEBs in all 167 neighboring stars within 2.5 arcminutes down to TESS-band magnitude 17.6, which covers an extra 0.5 magnitudes fainter than needed.

High Resolution Speckle Imaging

If an exoplanet host star has a spatially close companion, that companion (either gravitationally bound or coincidentally along the line of sight) can produce a false-positive transit signal if it is, for example, an eclipsing binary (EB). "Third-light flux" from the companion star can lead to an underestimated planetary radius if not accounted for in the transit model *(48)* and cause non-detections of small planets residing within the same exoplanetary system *(51)*. Close, bound companion stars, exist in almost one-half of FGK type stars *(52)*. To search for close-in bound companions unresolved in TESS or the ground-based follow-up observations, we obtained high-resolution imaging speckle imaging observations of GJ 367.

GJ 367 was observed on 2020 January 10 UT using the Zorro speckle instrument on Gemini South (*53, 54*). Zorro provides simultaneous speckle imaging in two bands (562 nm and 832 nm) with output data products including a reconstructed image and contrast limits on companions (*55*). Three sets of 1000 × 0.06 sec exposures were collected and subjected to Fourier analysis in a standard reduction pipeline (*56*). Figure S2 shows the contrast limits and the 832 nm reconstructed speckle image. We find that GJ 367 is a single star with no companion brighter than about 4-5 magnitudes below that of the target star (equivalent to an M5 dwarf) from the diffraction limit (20 mas) out to 1.2". At the distance of GJ 367 (d=9.4 pc) these angular limits correspond to spatial limits of 0.2 au to 11 au.

Stellar rotation period

WASP-South is an array of eight wide-field cameras, each equipped with 2k × 2k CCDs, that surveyed southern fields with a typical cadence of 15 mins (*57*). GJ 367 was observed over an interval of 150 nights in 2007 followed by 150 nights in 2008, at a time when WASP-South was equipped with 200-mm f/1.8 lenses observing with a broad, 400–700 nm bandpass. It was further observed over intervals of 160 nights in both 2013 and 2014, when WASP-South had 85-mm f/1.2 lenses with an SDSS r-band filter (*58*). We searched the accumulated data from each observing season for rotational modulations using established methods (*59*). Signals could arise from any star in the extraction aperture (48 arcsecs radius for the 200-mm lenses, 112 arcsecs for the 85-mm lenses), however GJ 367 is the brightest star in the aperture in both cases, with the next brightest star in the 200-mm aperture being 3 mag fainter.

For both 2007 and 2008 we find a modulation with a period of between 44 and 49 d, with a false-alarm probability <1 per cent in each case (Figure S3). The 2013 data instead show power at a longer period near 60 d (the reason for this is unknown, though it could result from phase changes caused by the emergence of star spots). The 2014 data show power at 24 d (compatible with being the first harmonic of the above period), plus power at the 44 – 49-d period itself. Combining all the data into one periodogram shows power between 44 and 49 d. We conclude that GJ 367 likely has a rotation period in that range, but cannot narrow it down further because a rotational modulation is incoherent over the data span. The amplitude of the modulation varies between 3 and 8 mmag.

A different methodology was also applied to the WASP data. We first calibrated the data as follows. We removed outliers if the difference between the magnitude of a data point and the median magnitude of the time series data is greater than 0.04. The data are re-binned to a cadence of 0.2 days and gaps inside each campaign are interpolated using the inpainting method based on the Multiscale discrete cosine transform (*60, 61*). We then apply a rotation pipeline that combines different methods to look for modulation. First we performed a time-period analysis using wavelets decomposition (*62, 63*). We also computed the auto-correlation function (ACF) that has been applied on large sample of stars (*64, 65*). Finally we compute the composite spectrum (CS), the product of the global wavelet power spectrum (that is the projection of the wavelet power spectrum on the period axis) with ACF. This CS allows us to emphasize signals present in both methods (*66, 67*). The results of this analysis are shown in Figure S4.

We find the excess of power around 48 days from different methods. The wavelet analysis shows more power at 23 days but we interpret this as the harmonic of the fundamental period. We therefore infer a rotation period of 48 ± 2 d, which agrees with the periodogram analysis described above. The TESS light curve for GJ 367 has an ambiguous variability (based on fast Fourier transform analysis, Lomb Scargle Periodogram, and the wavelet power spectrum), from which it is not possible to extract short periods with physical meaning (*68*).

The stellar rotation period can alternatively be estimated using the rotation-activity relationship for very low mass stars using the $R'_{HK}$ index obtained from the Ca II H and K emission lines (*69*). The log $R'_{HK}$ index of GJ 367 is -5.214 ± 0.074 (derived from the HARPS spectra described below) which translates to an estimate stellar rotation period of 58.0 ± 6.9 days. This is not inconsistent with the rotation period measured using the WASP data.

Spectroscopic observations

Twenty-four high-resolution (resolving power ~ 115000) spectra of GJ 367 were acquired during 24 different nights between 12 December 2003 and 07 February 2010 (UT) with the High Accuracy Radial velocity Planet Searcher (HARPS) spectrograph (*70*). The observations were carried out as part of a radial velocity (RV) survey (*71*). Shortly after the detection of the USP planet candidate in the TESS light curves, we acquired 81 additional HARPS spectra on 28 different nights between 23 June 2019 and 23 March 2020 (UT). In this second campaign we tailored our observing strategy to account for the short orbital period of GJ 367b by following a multi-visit approach, taking at least two spectra per night in 20 out of the 28 nights. The exposure times were set to 600 s and 900 s, except for two observations of 1500 s (06 June 2007, UT) and 1200 s (22 March 2020, UT) each, resulting in a S/N ratio per pixel at 650 nm between 3 and 127, with a median of 81. The exposure times used are less than 5% of the orbital period of the planet, limiting the smearing of any radial velocity signals. The spectra were automatically reduced using the HARPS Data Reduction Software (*72*). Two spectra (acquired at BJD$_{TDB}$: 2458665.548, 2458665.556) were rejected due to their low S/N ratio and they were not considered in the analysis. The phase coverage of the RV curve for GJ 367b is uneven (Figure 1) because the orbital period of the USP planet is close to a multiple of 24 hours. Therefore, observations taken over consecutive nights are at similar orbital phases. Subsequent additional follow-up observations were made to increase the RV phase coverage. However, this was limited by the visibility of the star and the observing facility shut down in March 2020 due to the Covid-19 pandemic.

In June 2015 the HARPS fiber bundle was upgraded to octagonal fibers (*73*). We therefore treated the HARPS RVs taken before and after June 2015 as two different data sets to account for the RV offset caused by the refurbishment of the instrument.

We extracted the RV measurements from the HARPS spectra, along with the Hα, Hβ, Hγ Balmer, Na D, and Ca H & K (S-index) lines, from which we computed activity indicators. The RVs were measured by maximizing the likelihood between each reduced spectra and a stellar template constructed from shifting all spectra to a common velocity frame and taking the median. The stellar template was shifted in several steps of RV in the range $27.93 - 67.61$ km s$^{-1}$ and for each step the likelihood was derived following established methods ($74$). The resulting RV time series has a standard deviation of 4.36 m s$^{-1}$ and a median precision of 0.74 m s$^{-1}$ in the range $0.45 - 9.05$ m s$^{-1}$. The RV measurements are listed in Table S1 and the corresponding activity indicators are listed in Table S2.

We also extracted additional spectral diagnostics, namely the chromatic index (CRX) and the differential line width (dLW), using the code SERVAL ($75$). These are shown in Figure S5.

### Stellar fundamental parameters

We determined the spectroscopic parameters of GJ 367 using the tool SPECMATCH-EMP ($76$) to compare the co-added HARPS spectrum with a spectral library of high-resolution (resolving power $\approx 55000$) spectra of FGKM stars. The tool is used to find the stellar effective temperature $T_{eff}$, logarithmic iron abundance [Fe/H], and stellar radius $R_s$. Since the spectra in the SPECMATCH-EMP library has a lower spectral resolution than the HARPS spectrum, we degrade the spectral resolution from ~115000 to ~60000 before applying SPECMATCH-EMP to the HARPS spectrum. Following the methods described in reference  We performed a Markov-Chain Monte Carlo simulation ($77$) which applies empirical relations ($78$) to derive the stellar mass $M_s$, stellar density $\rho_s$, surface gravity log $g$, and luminosity $L_s$ from the output stellar parameters of SPECMATCH-EMP and the distance $d$ from the Gaia parallax ($8$). There is a known systematic offset in the reported Gaia parallaxes, so we applied the recommended correction of +61 µas ($79$). The resulting properties of GJ 367 are listed in Table 1 and are used for the simultaneous light curve and RV analysis described below. As a cross-check, we derived the parameters of the host star using the Machine Learning tool ODUSSEAS ($80$). We used the co-added HARPS spectrum as input data to obtain $T_{eff} = 3405 \pm 70$ K and [Fe/H] = -0.04 $\pm$ 0.11 dex, consistent with the above analysis.

We also computed stellar parameters using the packages ISOCHRONES ($81$) and MIST ($82$) to fit models to the Gaia DR2 parallax ($\pi$) ($8, 83$) and 2MASS photometry (J,H,Ks bands) ($84$). We chose Gaussian priors on $T_{eff}$ and [Fe/H] based on our SPECMATCH-EMP results, and used MULTINEST ($85$) to obtain posterior samples. We obtained the following parameter estimates: $T_{eff} = 3674 \pm 35$ K, log $g = 4.820^{+0.007}_{-0.006}$ (cgs), [Fe=H] = 0.10 $\pm$ 0.09 dex, $M_s = 0.468^{+0.007}_{-0.006}$ M$_\odot$, $R_s = 0.440 \pm 0.004$ R$_\odot$, $\rho_s = 7.713 \pm 0.180$ g cm$^{-3}$, age = $8.0^{+3.8}_{-4.6}$ Gyr, distance $d = 9.410 \pm 0.005$ pc, and extinction A$_V = 0.07^{+0.09}_{-0.05}$ mag.

There are known discrepancies between observations and stellar evolution models of M dwarfs. In particular, the observed radii of these stars are larger than model predictions while stellar effective temperatures measured are often cooler ($85-88$). The stellar parameters derived from isochrone fitting may be less accurate, so we adopt the results obtained from the SPECMATCH-EMP tool in the subsequent joint analysis of the TESS photometry and HARPS RV.

The stellar chromospheric activity was determined from the Ca II H & K emission lines; the activity index has a median value of log ($R'_{HK}$) = -5.214 ± 0.074. The upper limit on GJ 367's X-ray flux is log($F_X$/mW m$^{-2}$)< -13.09 and bolometric flux of log($F_{bol}$/mW m$^{-2}$)=-7.94 mW m$^{-2}$ (90). The upper limit of the coronal X-ray activity of GJ 367 is thus log ($L_x/L_{bol}$) < 5.15, where $L_x$ is the X-ray luminosity and $L_{bol}$ is the bolometric luminosity.

The projected rotation velocity of GJ 367 has an upper limit of $v$ sin $i$ < 3 km s$^{-1}$ (90), equivalent to a rotation period of $P_{rot}$ > 7.71 days (assuming the system is aligned with the line of sight, i.e. $i$ = 90º. This agrees with the rotation periods estimated from the stellar rotation analysis of the WASP data, as well as the rotation period estimated from stellar chromospheric activity as described in the main text. Taking our the adopted rotation period of 48 ± 2 d, the rotational velocity of the star is $V_{rot}$ = 0.48 ± 0.02 km s$^{-1}$ (assuming $i$ = 90º). The semi-amplitude of the Rossiter-McLaughlin effect (91) is estimated to be 0.06 ± 0.01 m s$^{-1}$, which is small and negligible.

Using the Gaia DR2 parallax, proper motions, and RV, the Galactic space velocities of GJ 367 were found to be (U, V, W ) = (-11.732 ± 0.014, -36.53 ± 0.37, -21.928 ± 0.036) km s$^{-1}$. Based on established criteria (92), this gives the star a 97.6% probability of belonging to the thin disk, 2.4% of belonging to the thick disk, and a negligible probability of belonging to the Galactic halo. Thin disk membership is consistent with the value of [Fe/H]= -0.01 ± 0.12.

Table 1 lists the stellar parameters of GJ 367b and Table S3 lists its identifiers and additional photometric magnitudes of GJ 367b.

Stellar age

We used gyrochronology to estimate the age of GJ 367 following established methods [(93), their equation 32]. The global convective turnover timescale ($\tau_c$) for stars of mass 0.45 M$_\odot$ is $\tau_c$ = 1.769 × 10$^2$ d (94). We adopted the rotation period of 48 ± 2 d obtained from the WASP photometry. The estimated gyrochronological age of GJ 367 is then 3.98 ± 1.11 Gyr. This age is consistent with the isochronal age of 8.0$^{+3.8}_{-4.6}$ Gyr. Age-dating using gyrochronology may be imprecise for M dwarf stars. Stellar rotation studies have shown that stellar ages estimated from gyrochronology models may be under-predicted for large portions of the stars (65, 95). The photometric amplitudes of slowly rotating old stars are usually low and difficult to measure. As a result, gyrochronology models may be insufficiently calibrated due to the lack of old M dwarfs in the observed samples (94, 95). However, deriving ages of M dwarfs from isochrones may also be problematic because M dwarfs evolve slowly along the main-sequence and their observable properties do not change substantially. This means that isochrones in the Hertzsprung-Russell diagram are very close to one another across a range of ages, which makes estimating ages very difficult (*96*).

GJ 367 belongs to the Galactic thin disk, and its coronal X-ray and chromospheric indices indicate a low activity star. Along with the gyrochronology and isochrone placement considerations, GJ 367 is likely an older star with a lower age limit of 3.98 ± 1.11 Gyr.

Frequency analysis of RVs and activity indicators

We performed a frequency analysis of the HARPS RVs and activity indicators to search for the Doppler reflex motion induced by the transiting planet, additional orbiting companions and/or signals associated with stellar activity. We used only the 43 HARPS measurements taken in 2020 (after the fibre bundle upgrade) to avoid the presence of spurious peaks (aliases), such as those introduced by the yearly sampling, and to avoid having to account for an RV offset between the HARPS data-sets taken before and after the refurbishment of the instrument.

We show in Figure S5 the generalized Lomb-Scargle periodograms (GLS) (*97*) of the 2020 HARPS measurements and activity indicators for two frequency ranges encompassing the orbital frequency of the USP planet and the frequency at which we expect to see the stellar signal. False alarm probabilities (FAPs) at 0.1 %, 1 %, and 5 %, were derived using a bootstrap method (*98*). The periodograms of the HARPS RVs and Hα, Hβ, dLW activity indicators show their strongest power at frequencies lower than 0.03 $d^{-1}$, with the periodograms of the RVs having its highest peak at $f_1 = 0.028$ $d^{-1}$ (~36 days). Given the 38-day baseline of the 2020's HARPS measurements, this low-frequency signal is compatible with the rotational modulation detected in the periodogram of the WASP-South data and it is very likely associated with the presence of active regions (spots and plage) carried around by stellar rotation. Following a pre-whitening technique (*99*), we fitted sine models to the amplitude and phase at the first dominant frequency ($f_1$) and subtracted the model from the time series. The GLS periodogram of the RV residuals (Fig. S5B) shows a peak at $f_2 = 0.085$ $d^{-1}$, corresponding to a period of ~11.7 days, which is also visible in the periodogram of the RVs and not seen in the periodograms of the activity indicators. Although the peak is not statistically significant (FAP ≈ 1 %), it might be induced by the presence of an additional planet in the system with a mass 5.7 $M_\oplus$ (3-sigma upper limit). We iterated by removing the signal at $f_2$. The periodogram of the RV residuals, following the subtraction of $f_1$ and $f_2$, shows its strongest power at the orbital frequency of the transiting planet $f_3 = 3.106$ $d^{-1}$ (Fig. S5C). The same peak is not seen in any of the periodograms of the activity indicators, suggesting that the signal at $f_3$ is induced by the orbital motion of GJ 367b around its host star.

To assess the significance of the Doppler signal of GJ 367b, we used an implementation (*100, 101*) of the floating-chunk offset (FCO) method. For USP planets, within a given night the observed RV variation is mainly induced by the orbital motion of the close-in planet. We assume that additional longer period signals, such as those due to stellar rotation, magnetic cycles, and additional outer planets, remain constant within a given night, introducing a nightly offset that changes from night to night. If we can sample a sufficient fraction of the orbit, then these nightly chunks can be shifted until the best fitting model of the orbital motion of the star is found.

We used the FCO method as a periodogram and searched for the stellar Doppler reflex motion induced by the USP planet. For this analysis, we used the HARPS RVs taken in 2019-2020, as most of these RV measurements were acquired following a nightly multi-visit strategy to cover a large fraction of the orbit of the transiting USP planet within each observing night. We divided the 2019-2020's HARPS RVs into subsets of nightly measurements and analyzed only those subsets that contain multiple measurements per night, leading to a total of 20 chunks. For each trial frequency within the 0.5–6.0 d$^{-1}$ range with a resolution of 0.0001 d$^{-1}$, we fitted the subsets of HARPS data using a sine function model, while free parameters of the RV semi-amplitude, phase, and 20 nightly offsets. Figure S6 displays the reduced $\chi^2$ as a function of each trial frequency. The minimum is found at the orbital frequency of the transiting planet ($f_b$ = 3.106 d$^{-1}$), with an RV semi-amplitude variation of $0.76 \pm 0.18$ m s$^{-1}$, confirming the detection of the transiting planet in our HARPS data.

Simultaneous analysis of the TESS photometry and HARPS RV measurements

The SAP light curve was used for the joint analysis of GJ 367. After removing non-arithmetical values either in the time, flux or flux-uncertainty record, 16,535 flux measurements remained. These measurements and their uncertainties were normalized by dividing all of them by their median value.

The light curve was modelled using the Transit and Light Curve Modeller (TLCM) *(13)* code, which performs joint RV and light curve fitting with a wavelet-based red noise model *(102)*. It uses two parameters, the white noise level $\sigma_w$ and the red noise factor $\sigma_r$, to characterize the noise in the light curve. The wavelet parameters are fitted alongside the free parameters of the modelled system. A prior is applied when fitting the wavelet parameters such that the 1-sigma scatter of the light curve residuals is equal to the average uncertainties of the photometric data, to avoid over-fitting the data. Transit injection analysis using the short cadence Kepler SAP light curves shows that the wavelet-based method can remove stellar variability and instrumental effects, and recover the planet-to-star-radius-ratio with uncertainty better than 15% for cases where the signal-to-noise ratio of a transit is low *(103)*. This planet-to-star-radius-ratio uncertainty reduces with increasing number of transits and increasing S/N ratio.

A time resolution parameter is used in the TLCM model which determines the number of sub-exposures that are used to integrate the modelled flux over time. This smoothed light curve model is compared to the observed data to calculate the goodness of fit. This is required when the exposure time is long compared to the length of ingress/egress of a transit, which is the case for GJ 367 b. In this case, we used a time resolution parameter of 5 in our model. Two points from the RV data set were removed because they were obtained during twilight and their uncertainties became large (15 and 77 m s$^{-1}$, respectively). The FCO method *(100, 104)*, as described above, was applied to fit the RV curve where a nightly RV offset value for the nightly groups of RV measurements. This method requires that at least two or more RV points to be obtained over the same night in each RV subset. From the available remaining 103 RV points we found that 73 can be used for the analysis following the FCO method. We used 19 nightly RV offsets for the twenty nights, and free parameters of the systemic velocity $V_\gamma$, the RV amplitude $K$. The tidal circularization timescale is 21 Myr (see below), so the orbit of GJ 367b is likely to be circular given the host star age. We adopted a circular orbit in our model. The additional free parameters transit epoch $T_0$ and the period $P$ are shared with the photometric data. The photometry was characterized by the scaled semi major axis $a/R_s$, the planet-to-star radius ratio $R_p/R_s$, the impact parameter $b_c^2$, a flux zero point shift $p_0$, and two parameters for the wavelet-based red noise analysis: $\sigma_r$ and $\sigma_w$, the red-noise factor and the white noise level *(102)*. Contaminating light from the neighbouring stars estimated from ground-based observations was also taken into account, where a third light contamination parameter, $l_3$, is fitted to correct for the flux dilution in the light curve model. We used a penalty function when the standard deviation of the (red noise corrected) light curve residual was not equal to the mean of the photometric uncertainties *(13)*.

For the joint light curve, RV and FCO fitting there were a total of 19 RV offsets + 2 RV parameters + Epoch, Period + 7 light curve parameters = 29 free parameters. During the joint fitting the limb darkening coefficients (u+ and u-) were fixed at u+ = 0.62 and u- = 0.15 *(105)*. We used TLCM to search for an initial solution using the Harmony Search type of Genetic Algorithms, refined by Simulated Annealing *(13)*. The estimation of final parameters and their uncertainties were done via an MCMC analysis using 10 independent chains and one million steps in each chain with a thinning factor of 100. The convergence was estimated by the Gelman-Rubin statistic *(106)* and we prescribed that the estimated sample size *(107)* should be at least 200 before stopping the MCMC analysis. The results of the joint model are listed in Table 1 and the fit to the data is shown in Figure 1. The corner plots showing the posterior probability distributions are shown in Figure S7.

In the joint fit, the stellar radius and the log $g$ values were used as priors. We obtained the stellar effective temperature, metallicity and log $g$ from the stellar spectral analysis (see above). The mean stellar density is determined from the period and the scaled scaled semi-major axis $a/R_s$ via Kepler's third law *(13, 108, 109)*. Matching the stellar effective temperature, metallicity and the mean stellar density with isochrones, we obtain a stellar radius $R_s = 0.459 \pm 0.003\ R_\odot$ and mass $M_s = 0.500 \pm 0.005\ M_\odot$ which is consistent to the values obtained from SPECMATCH-EMP analysis or via the Gaia-parallax value. We used the log $g$ and the stellar radius value as Gaussian distributed priors (Table 1).

To check for consistency, we performed a Gaussian Process (GP) analysis of all available RV data following a modified version of existing methods *(110, 111)*. The GP models the correlated noise in the RV data arising from surface inhomogeneities of the host star to disentangle the planetary signal in the dataset *(110)*. We used the TESS light curve to constrain the hyperparameters of the GP kernel. We could not detect the suspected 40-day stellar rotation period in the TESS light curve, we therefore impose a broad log-uniform prior (1–200 days) on the periodicity of the GP model. The observed flux variation in the TESS light curve does suggest a short correlation timescale of about 3 days. This could be caused by spacecraft instrumental effects or detrending in the PDC pipeline. If we trust this correlation timescale is astrophysical, our GP analysis showed a K-amplitude of $0.87^{+0.16}_{-0.15}$ m s$^{-1}$, which is consistent with the FCO method. Given the uncertainty in the GP model, we adopted the FCO value as our fiducial result. As a comparison, we also used TLCM to perform a combined RV and TESS light curve analysis without the FCO method, which made use of all available RV data. This yielded a RV amplitude of 1.19 ± 0.21 m s$^{-1}$, corresponding to a planet mass of 0.79 ± 0.14 M$_\oplus$ and a planet density of 18.33 ± 13.36 g cm$^{-3}$. Without accounting for the RV variation induced by stellar variation or possible long period companions, the amplitude of the RV curve is inflated as expected, which resulted in a planet density which we regard as unphysically high.

<u>Interior structure of the USP planet</u>

The interior composition of GJ 367b was determined with a machine learning approach *(32)* which utilizes the multitask *(112)* mixture density networks (MDNs; *(113)*) to predict the interiors of planets. An MDN is trained using a large dataset of synthetic planets assumed to consist of four distinctive layers: iron core, a silicate mantle, ice shell, and H/He envelope). The set of synthetic planets with random interior structures is used to train the MDN. For a set of input planetary mass and radius, the trained MDN predicts the full conditional probability distribution which can then infer the relative thickness of each interior layer (*113*). The resulting distributions depend on the assumed prior distributions, as it is the case for other inference methods *(114)*. We adopt priors which we consider very conservative, each layer having a linear prior distribution between 0 and 1. The effect of a different prior for the light gas layer (e.g. logarithmic sampling instead of linear), has minor effects on the resulting distributions, especially for the core and silicate layers *(32)*, which dominate the interior structure of GJ 367b. The core and silicate layers use standard equations of state (EoS) *(115)*, the ice layer is modelled using the EoS of water ice VII. Liquid or superionic phases are not taken into account, which would increase the degeneracy of the solutions. The gas layer is modelled with a zero-temperature Thomas-Fermi-Dirac EoS (*116*). For a highly-irradiated USP planet, we expect this EoS to overestimate the density of the layer. Thus, the inferred thickness for the gas layer in Figures 3, S8, and S9 can be considered as upper bounds. Ice and gas could be present if the object formed far from the parent star and reached its current position via migration *(117)* or eccentricity excitation *(118)*. This method has been used *(32)* to compare the inferred interior structure of the Earth to the core and mantle thickness from the Preliminary Reference Earth Model (PREM) *(119)*. The values between two models agree within the uncertainties but the peak of the MDN distribution does not align with the core values from PREM, which was attributed to the large degeneracy in the interior model. This degeneracy can only be reduced if there are additional planet interior constraints.

We used the best-fitting mass and radius of GJ 367b from the joint analysis as input parameters to evaluate the planet's bulk composition. The predicted interior composition of GJ 367b is shown in Figure 3 where the median relative thickness of the planet's iron core is $86 \pm 5\%$. Figure S9 shows the predicted interior compositions of Earth, Mars and Mercury. In the case of Mercury, the predicted iron core radius fraction is $81 \pm 4\%$, consistent with the iron core radius fraction measured by the MESSENGER spacecraft (*30*). This implies that both interiors of GJ 367b and Mercury are predominately composed of iron, so may have formed via similar mechanisms. The density of GJ 367b could be as low as 5.941 g cm⁻³ and as high as 10.271 g cm⁻³ due to the uncertainty on the planetary mass and radius. We modelled the interior structure of the upper and lower density cases using MDN. In both limiting cases (Figure S8), the iron radius fraction of GJ 367b is larger than the Earth's.

**Supplementary Text**

Tidal evolution timescales
The tidal evolution timescale $\tau_a$ *(120, 121)* is defined as

$$\tau_a = \frac{P}{\dot{P}} = -\frac{2PQ_s}{27\pi}\left(\frac{M_s}{M_p}\right)\left(\frac{a}{R_s}\right)^5, \qquad (S1)$$

where $P$ is the orbital period, $\dot{P}$ is the rate of change of the orbital period, $Q_s$ is the tidal quality factor of the star which has a value in the typical range of $10^{6-9}$, and $a$ is the semi-major axis. The minimum evolution timescale can be obtained by assuming $Q_s = 10^6$. In such case, we find that it would require at least 2.88 Gyr for the orbit of GJ 367b to shrink to zero based on the current orbital decay rate.

The tidal circularization timescale *(120, 121)* is expressed as

$$\tau_e = \frac{e}{|de/dt|} = \frac{2Q_p}{63\pi}\left(\frac{M_p}{M_s}\right)\left(\frac{a}{R_s}\right)^5 P, \qquad (S2)$$

where $e$ is the eccentricity, $|de/dt|$ is the rate of change of the eccentricity, $Q_p$ is the planet quality factor. We find that $\tau_e \sim 21$ Myr if $Q_p$ of GJ 367 is assumed to be $10^6$.

Mass-Radius diagram data selection
The planetary system data were obtained from the NASA Exoplanet Archive *(122)*. We downloaded the planetary system table on 26 August 2021 and selected planets with radii smaller than 2 $R_\oplus$ and masses lower than 20 $M_\oplus$. We further selected planets which have mass and radius precisions better than 30%. Sixty-five planets met these requirements, including GJ 367b. The masses and radii of rocky planets are shown in Figure 2.

<u>Transmission and thermal emission metrics</u>

We calculate the transmission and thermal emission metric *(123)* for GJ 367b to evaluate the suitability of the planet for atmospheric characterisation. The transmission metric (TSM) is defined as:

$$TSM = 0.19 \times \frac{R_p^3 T_{eq}}{M_p R_s^2} \times 10^{-m_J/5}. \qquad (S3)$$

$R_p$, $M_p$, $R_s$ are in units of Earth radii, Earth masses and Solar Radii, respectively. For planets with unknown masses, their masses can be estimated using an empirical mass-radius relation *(124)* [*(123)*, their equation 2]. $T_{eq}$ is the equilibrium temperature of the planet assuming zero albedo and $m_J$ is the apparent magnitude of the host star in the J band. The emission spectroscopy metric (ESM) is defined as:

$$ESM = 4.29 \times 10^6 \frac{B_{7.5}(T_{day})}{B_{7.5}(T_{eff})} \times \left(\frac{R_p}{R_s}\right)^2 \times 10^{-m_K/5}. \qquad (S4)$$

The Planck function, $B_{7.5}$, is evaluated for a given temperature at a representative wavelength of 7.5 μm, the day-side temperature $T_{day}$ is estimated as $1.10 \times T_{eq}$, and $m_K$ is the apparent magnitude of the host star in the K band.

Using the above metrics, we find that GJ 367b has a TSM of 41.9 and an ESM of 16.1. This places GJ 367b well above the threshold values for both transmission (TSM = 10) and emission (ESM = 7.5) spectroscopy of terrestrial planets ($R_p < 1.5\ R_\oplus$) *(123, 125)*. Figure S10 compares the TSM and ESM of GJ 367b with known terrestrial planets.

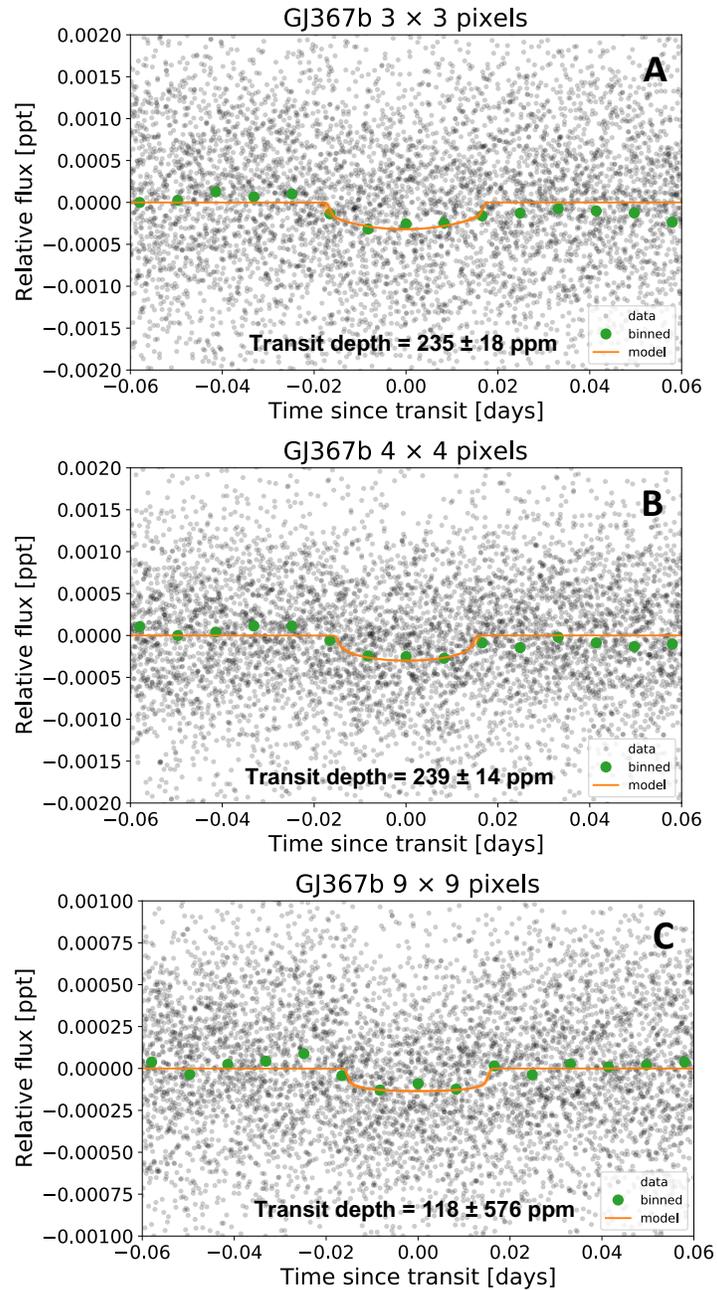

**Figure S1. Phase-folded transit light curves of GJ 367 extracted using different apertures.** **(A)** Transit light curve of the 3 × 3 pixels aperture. **(B)** Transit light curves of the 4 × 4 pixels aperture. **(C)** Transit light curves of the 9 × 9 pixels aperture. The measured transit depths and aperture sizes are not correlated which implies that the transit signal is originated from GJ 367.

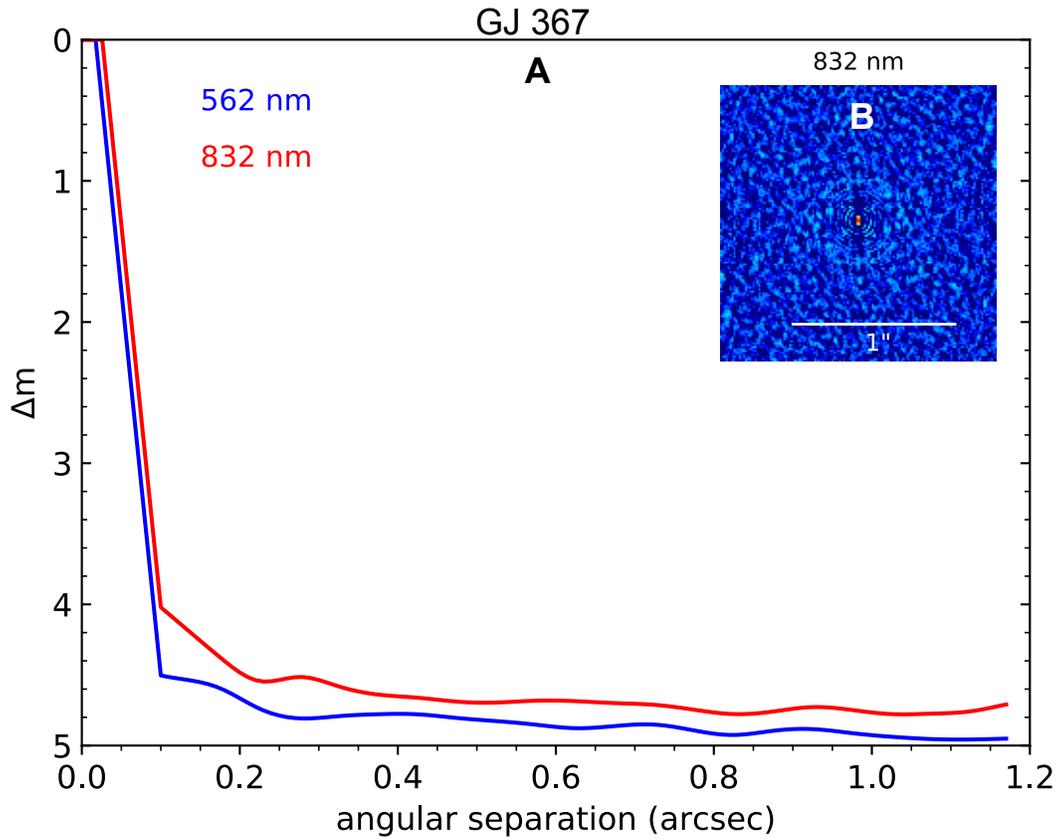

**Fig. S2. Contrasts limits and speckle imaging of GJ 367. (A)** Limits obtained from speckle imaging at 'Zorro 562 nm speckle imaging (blue), and 'Zorro 832 nm speckle imaging (red) are shown as functions of angular separation up to 1.2''. Δm is the magnitude contrast. **(B)** Zorro speckle imaging observed in the 832 nm filter. No companion brighter than 4-5 magnitudes below that of GJ 367 are detected out to 1.2''.

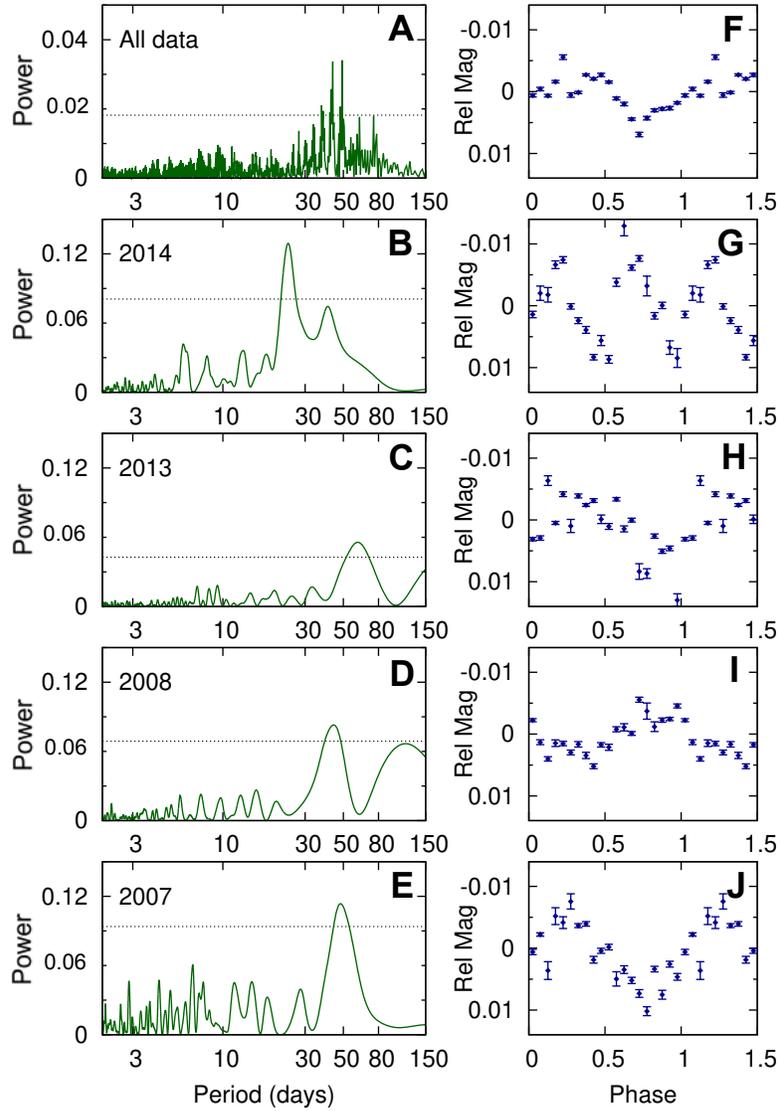

**Fig. S3. Lomb-Scargle periodogram analysis of WASP photometry.** Periodograms computed using data from **(A)** all WASP data and individual seasons, **(B)** 2014, **(C)** 2013, **(D)** 2008 and **(E)** 2007. The horizontal lines in panels **(A)** to **(E)** indicate the estimated 1% false-alarm level. Panels **(F)** to **(J)** shows the phase-folded and binned light curves using the maximum peak periods in the corresponding periodograms.

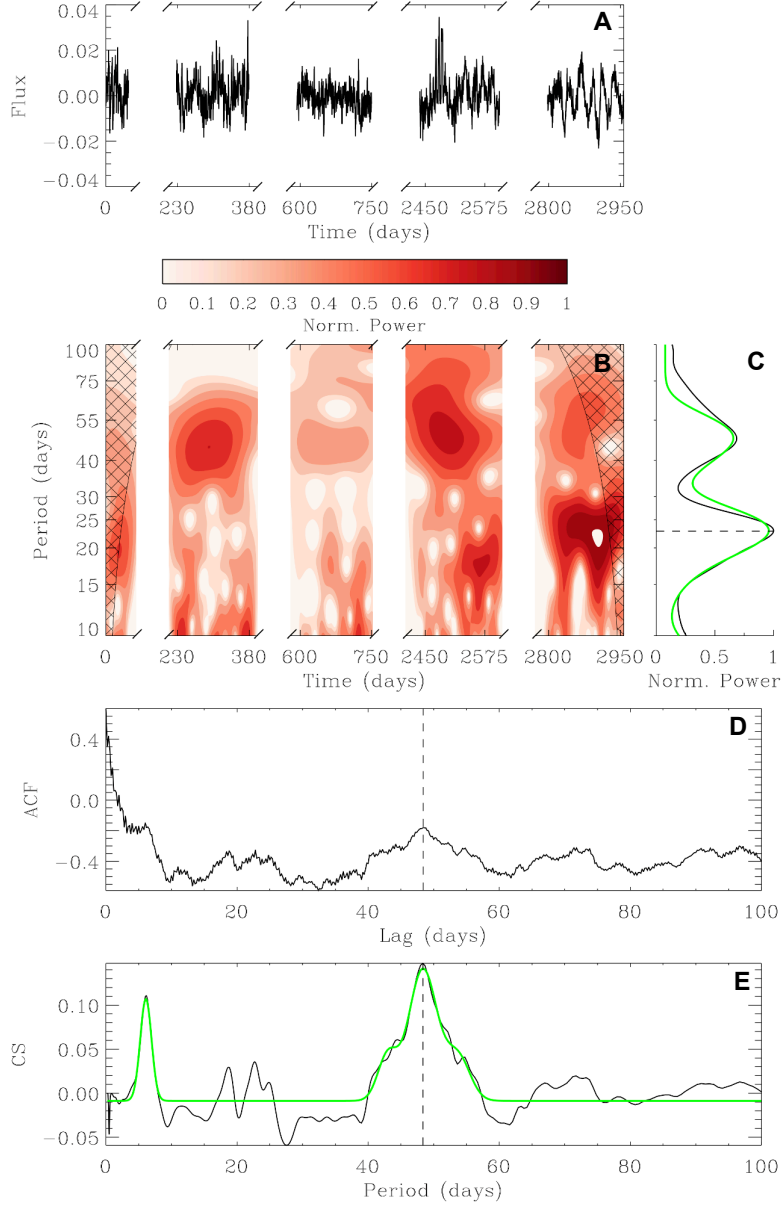

**Fig. S4. Rotation period analysis of the WASP time series data.** The time zero point is Heliocentric Julian Date (HJD) 2453860. **(A)** The time series WASP photometry. **(B)** The-period analysis of the WASP data. In order to measure a rotation period with the wavelet analysis, a minimum length of the time series is needed in order to search for a given Prot. Therefore, hatched regions on either end of the figure indicate the "cone of influence" where edge effects become too important to retrieve reliable $P_{rot}$ (*62*). **(C)** The projection of the wavelet power spectrum on the period axis (GWPS). The horizontal dashed line correspond to the highest peak fitted in the wavelet analysis. **(D)** The Auto-correlation function analysis. The vertical dashed lines correspond to the highest peak fitted in the analysis (that could be the rotation period or one of its harmonics) **(E)** The Composite Spectrum of ACF and Wavelet analyses. The green lines in the CS and the GWPS are the fitted Gaussian functions and the vertical dashed line indicates the highest peak fitted in the analysis.

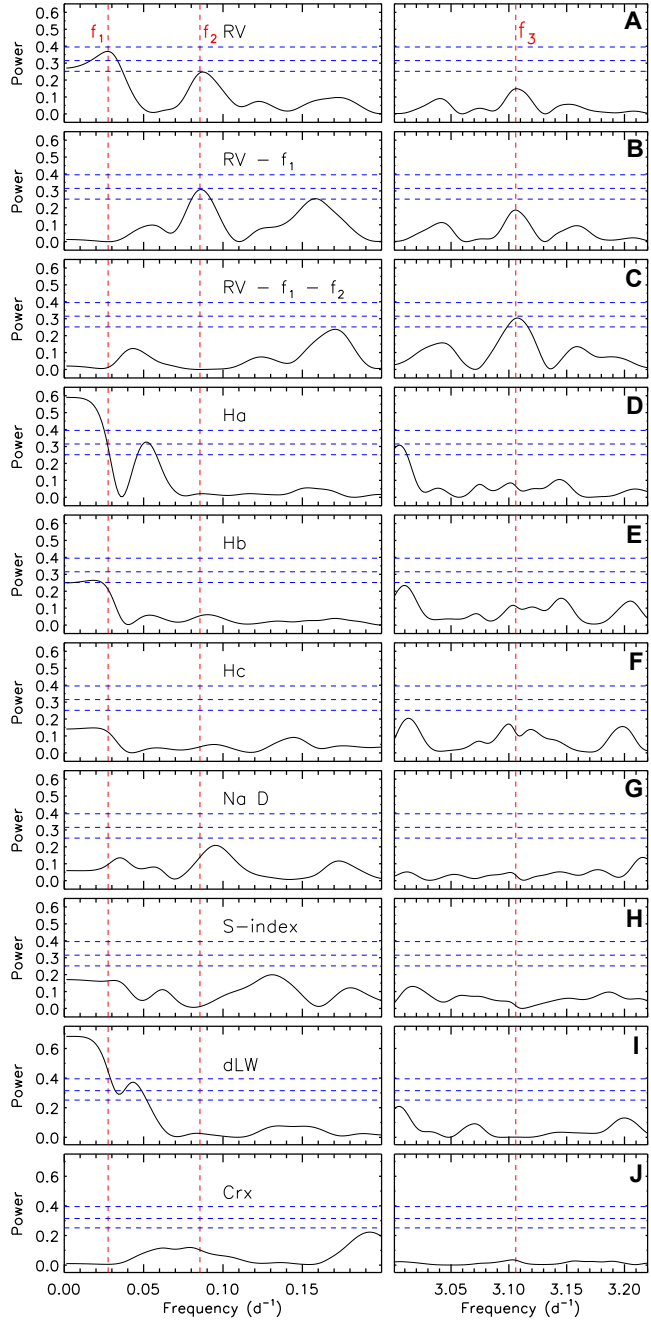

**Fig. S5. Generalized Lomb-Scargle periodograms of the 2020 HARPS RVs and activity indicators encompassing two frequency ranges (right and left panels). (A)**: RVs; **(B)** RV residuals following the subtraction of the signal at $f_1 = 0.028$ d⁻¹; **(C)** RV residuals following the subtraction of the signals at $f_1 = 0.028$ d⁻¹ and $f_2 = 0.085$ d⁻¹; **(D)** Hα; **(E)** Hβ; **(F)** Hγ; **(G)** Na D; **(H)** S-index; **(I)** differential line width (dLW); **(J)** chromaticity (Crx). The vertical dashed red lines mark the positions of the frequencies at $f_1$, $f_2$, and $f_3$ as labelled, with the latter being the orbital frequency of GJ 367b ($f_3 = 3.106$ d⁻¹). The horizontal dashed blue lines mark the FAPs of 0.1, 1, and 5 % (from top to bottom), estimated using a boot-strap method *(98)*. FAPs were computed for a blind search of periodicities in the RV data set, without prior information from the TESS transits.

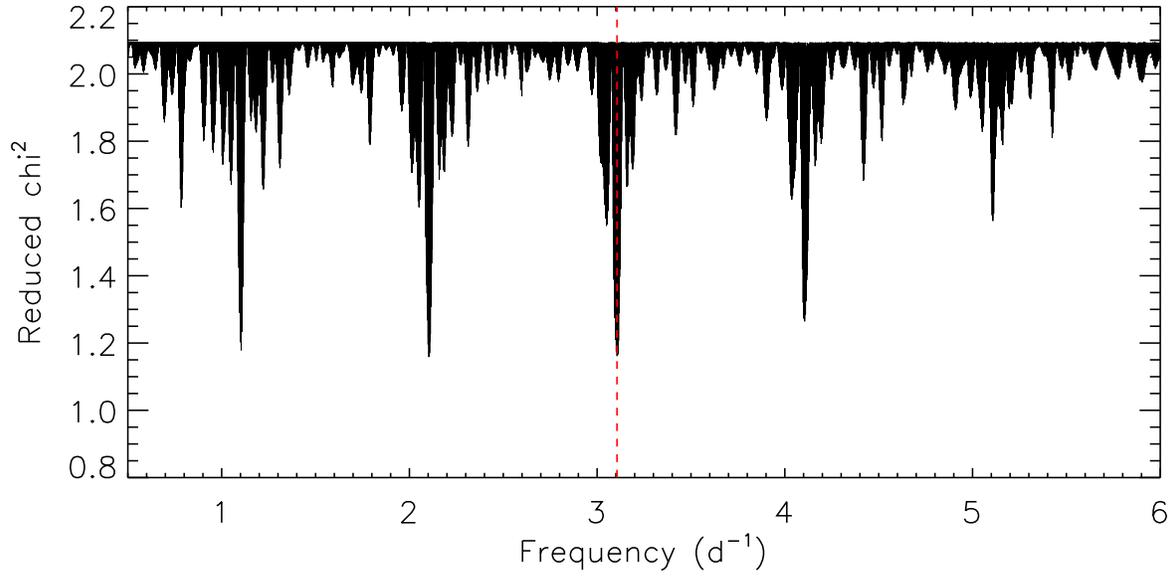

**Fig. S6. Floating chunk offset periodogram of the 20 subsets of HARPS data that include multiple observations per night.** The red dashed vertical line marks the orbital frequency of GJ 367b. The equally spaced minima separated by 1 d$^{-1}$ are the 1-day aliases of the orbital frequency.

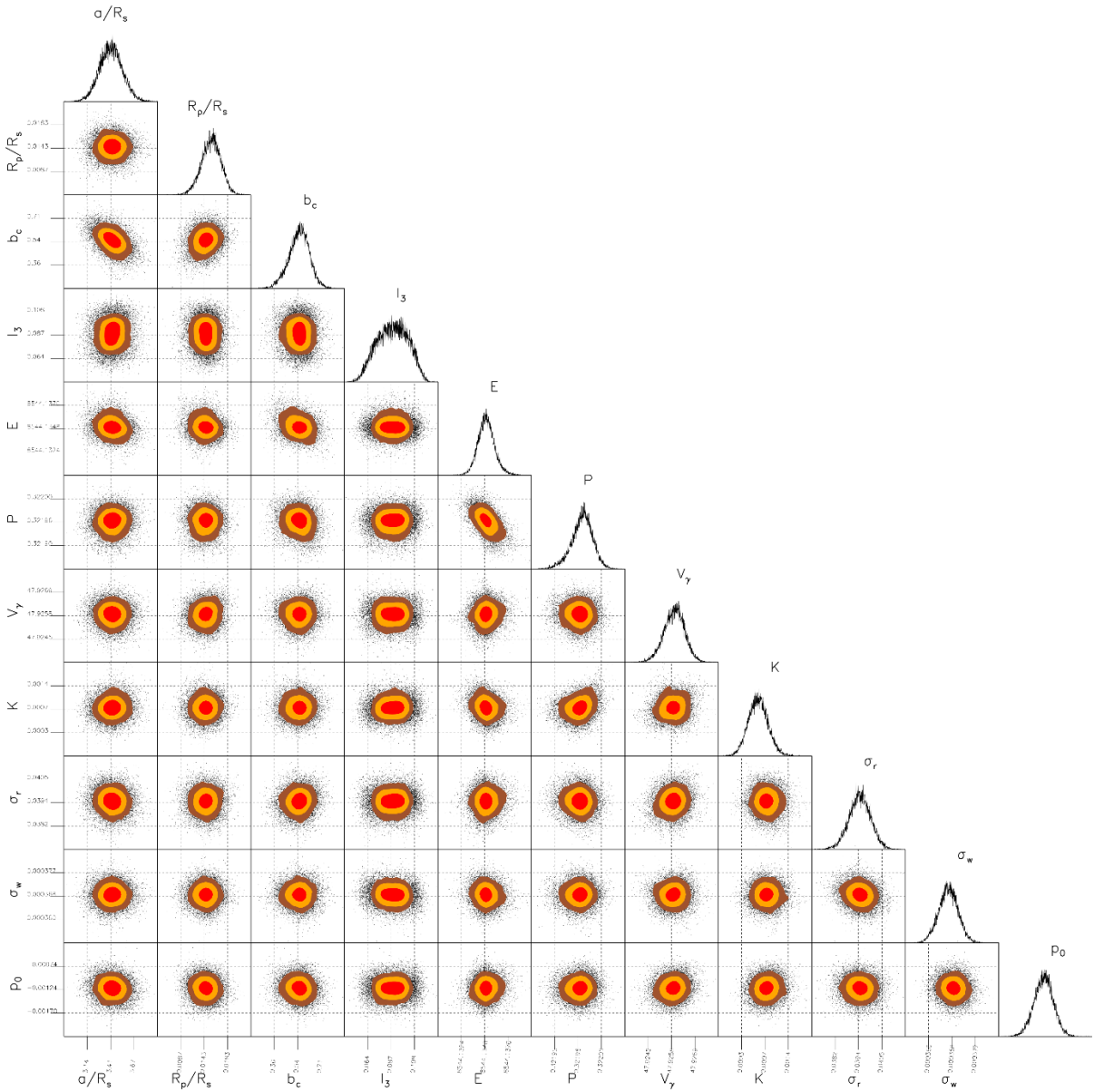

**Fig. S7. Corner plot showing the posterior probability distributions of the model parameters.** The model was fitted using the joint light curve, RV and FCO fitting method described above. The plot compares the posterior distributions of fitted parameters on the x-axis against each other on the y-axis. The contours in each panel denoted the 1- (red), 2- (orange), and 3- (brown) sigma levels. The projected posterior distributions of each parameter are shown on the top of each column.

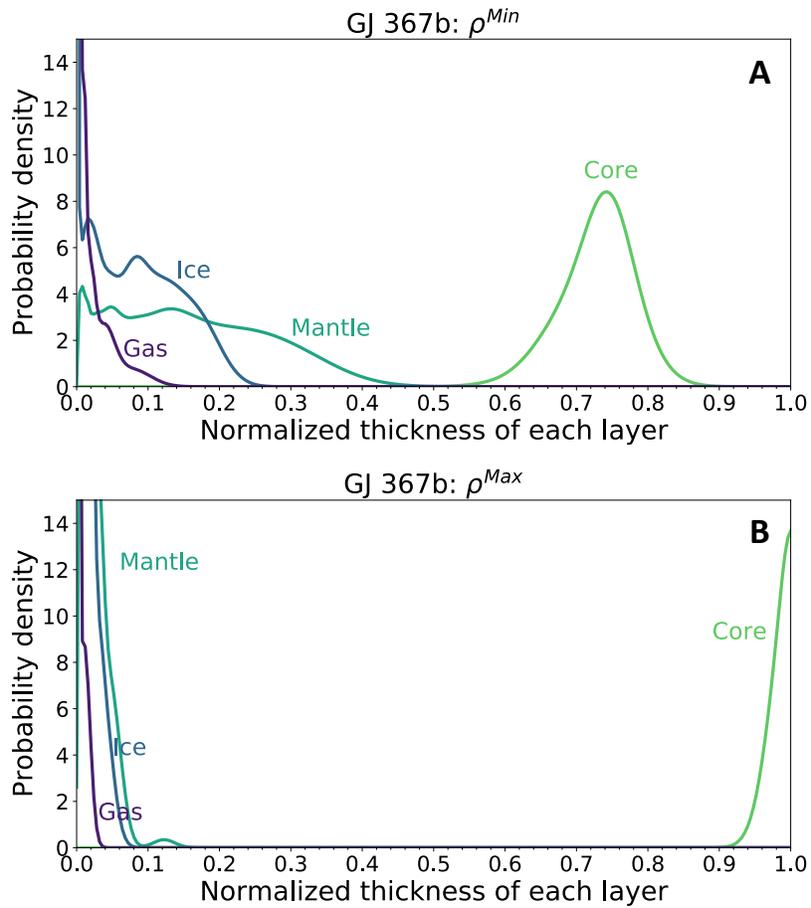

**Fig. S8. Extreme range of interior models GJ 367b.** Same as Figure 3B but using the lower ($\rho^{\text{Min}}$; top panel) and upper ($\rho^{\text{Max}}$; bottom panel) density limits inferred from the uncertainties on the planetary mass and radius measurements. The iron radius fraction of the lower density limit case is $78 \pm 7\%$ and the iron radius fraction of the upper density limit is $98.1 \pm 2.8\%$. In both cases, the iron radius fractions are larger than the Earth's model prediction (Figure S9).

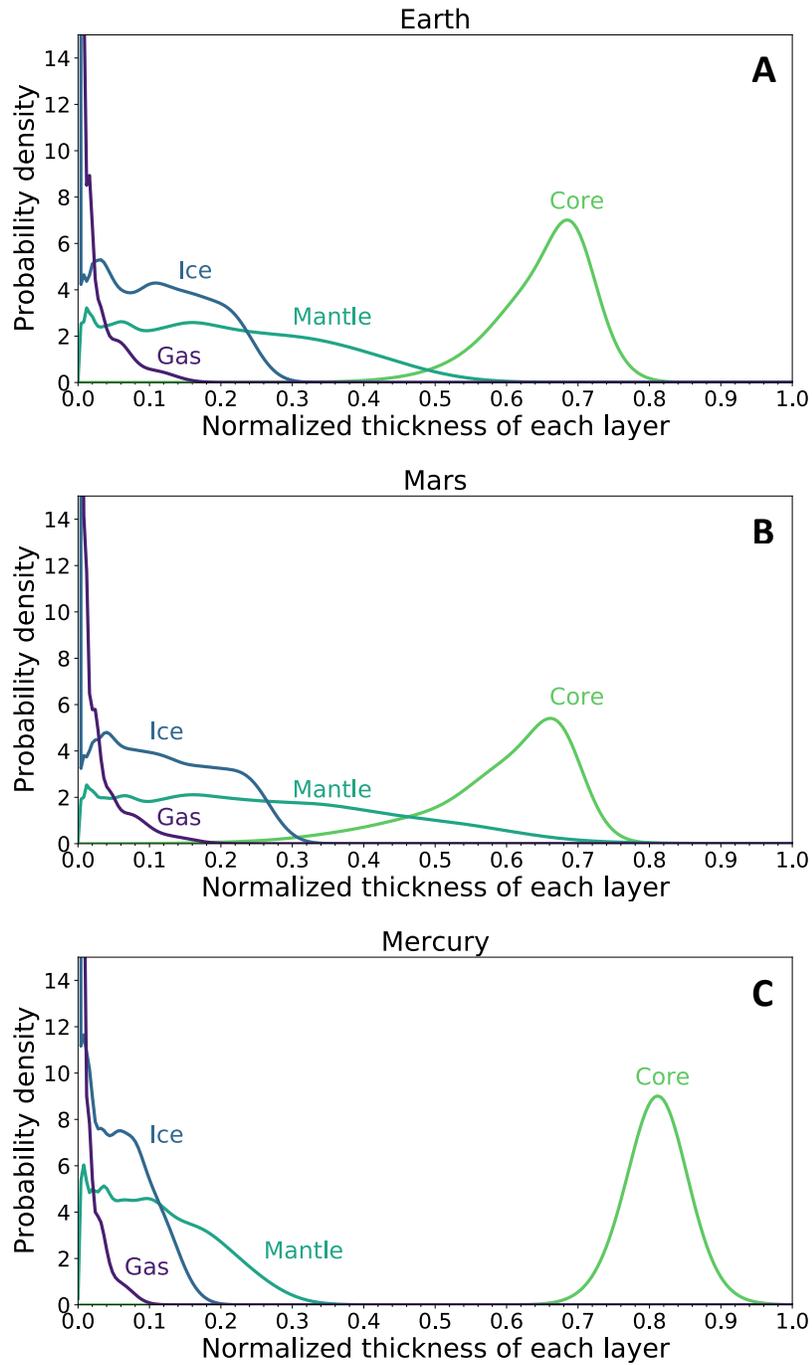

**Fig. S9. Interior model for Solar System terrestrial planet**. Same as Figure 3B but for **(A)** Earth, **(B)** Mars, and **(C)** Mercury, using only mass and radius as observable parameters. The relative thicknesses of the interior layers of GJ 367b (Figure 3) are consistent with the MDN prediction for Mercury (iron fraction = 81.1 ± 4.4%) as well as the measured value from the MESSENGER spacecraft (*30*).

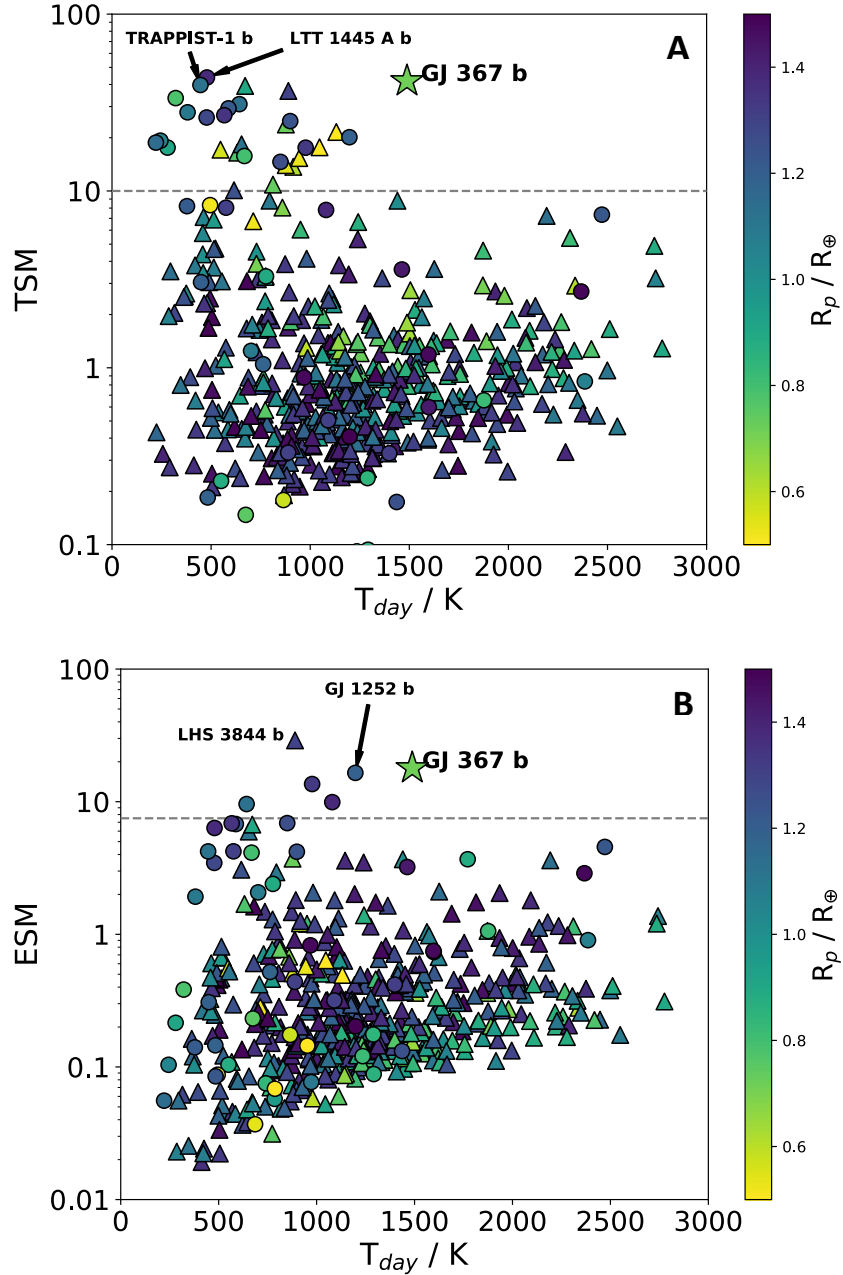

**Fig. S10. Transmission and thermal emission metrics of known terrestrial planets with $R_p < 1.5$ $R_⊕$. (A)** Transmission metric, TSM. **(B)** Thermal emission metric, ESM. In both panels, planets directly measured masses are denoted by circles. Triangles denote planets with masses estimates from an empirical mass-radius relation *(124)*. The TSM and ESM of GJ 367b are indicated by the star symbol. The color of each point indicates the size of the planet relative to Earth. The threshold TSM and ESM values are denoted by the grey lines. Because GJ 367b lies above these lines, it is suitable atmospheric studies. Selected other planets are labelled.

**Table S1. Time series radial velocities from HARPS spectra.** RV measurements obtained before and after the instrument upgrade are denoted HARPS and HARPS+, respectively. A machine-readable version of the table is available as Data S1.

| Time<br>BJD - 2450000 | RV<br>[m s$^{-1}$] | RV<br>uncertainty<br>[m s$^{-1}$] | Instrument |
|---|---|---|---|
| 2985.779966 | 47935.28 | 1.19 | HARPS |
| 2996.800243 | 47931.50 | 0.62 | HARPS |
| 3366.861538 | 47933.22 | 0.51 | HARPS |
| 3369.821915 | 47937.17 | 0.47 | HARPS |
| 3370.838755 | 47940.01 | 0.66 | HARPS |
| 3371.798971 | 47938.15 | 0.53 | HARPS |
| 3373.834165 | 47937.76 | 0.65 | HARPS |
| 3374.821349 | 47936.35 | 0.56 | HARPS |
| 3375.794875 | 47935.50 | 0.50 | HARPS |
| 3376.803020 | 47936.20 | 0.55 | HARPS |
| 3378.725045 | 47935.75 | 0.51 | HARPS |
| 3491.542478 | 47936.40 | 1.03 | HARPS |
| 4120.788138 | 47937.95 | 0.45 | HARPS |
| 4134.749658 | 47937.21 | 0.58 | HARPS |
| 4199.598451 | 47940.60 | 0.60 | HARPS |
| 4231.581759 | 47938.80 | 0.70 | HARPS |
| 4257.511359 | 47934.97 | 0.62 | HARPS |
| 4569.564214 | 47938.62 | 0.55 | HARPS |
| 4919.604700 | 47939.40 | 0.54 | HARPS |
| 4935.546236 | 47927.66 | 0.81 | HARPS |
| 4937.573143 | 47932.03 | 0.66 | HARPS |
| 4993.478922 | 47936.89 | 0.68 | HARPS |
| 5233.697882 | 47938.58 | 0.69 | HARPS |
| 5234.675358 | 47937.51 | 0.99 | HARPS |
| 8658.453921 | 47928.04 | 0.82 | HARPS+ |
| 8658.461142 | 47929.94 | 0.81 | HARPS+ |
| 8658.468654 | 47926.79 | 0.84 | HARPS+ |
| 8658.475609 | 47927.62 | 0.78 | HARPS+ |
| 8658.483201 | 47930.32 | 0.88 | HARPS+ |
| 8658.490238 | 47929.15 | 0.85 | HARPS+ |
| 8658.497610 | 47929.19 | 0.84 | HARPS+ |
| 8658.504844 | 47928.82 | 0.89 | HARPS+ |
| 8658.512158 | 47928.89 | 0.87 | HARPS+ |
| 8658.519681 | 47926.86 | 0.82 | HARPS+ |
| 8658.526845 | 47927.70 | 0.85 | HARPS+ |
| 8658.534217 | 47928.82 | 0.81 | HARPS+ |
| 8658.541242 | 47928.86 | 0.84 | HARPS+ |

| Time BJD - 2450000 | RV [m s⁻¹] | RV uncertainty [m s⁻¹] | Instrument |
|---|---|---|---|
| 8665.468148 | 47929.06 | 1.05 | HARPS+ |
| 8665.475520 | 47929.82 | 1.06 | HARPS+ |
| 8665.482893 | 47927.33 | 1.00 | HARPS+ |
| 8665.489223 | 47927.95 | 1.60 | HARPS+ |
| 8665.498748 | 47928.08 | 1.81 | HARPS+ |
| 8665.504107 | 47929.38 | 1.41 | HARPS+ |
| 8665.512937 | 47930.01 | 1.85 | HARPS+ |
| 8665.518712 | 47927.33 | 1.97 | HARPS+ |
| 8665.526779 | 47925.91 | 2.80 | HARPS+ |
| 8665.533526 | 47926.95 | 3.27 | HARPS+ |
| 8665.541107 | 47930.57 | 9.05 | HARPS+ |
| 8665.548063 | 47913.66 | 15.86 | HARPS+ |
| 8665.555852 | 47866.39 | 77.67 | HARPS+ |
| 8666.447238 | 47930.52 | 0.76 | HARPS+ |
| 8666.454414 | 47931.84 | 0.86 | HARPS+ |
| 8666.461716 | 47929.85 | 0.87 | HARPS+ |
| 8666.468811 | 47932.17 | 0.98 | HARPS+ |
| 8666.476253 | 47929.61 | 0.99 | HARPS+ |
| 8666.483764 | 47929.98 | 1.12 | HARPS+ |
| 8666.491136 | 47931.60 | 1.09 | HARPS+ |
| 8666.498300 | 47930.22 | 1.07 | HARPS+ |
| 8666.505464 | 47930.01 | 1.23 | HARPS+ |
| 8666.512558 | 47928.10 | 1.31 | HARPS+ |
| 8666.520625 | 47930.41 | 1.36 | HARPS+ |
| 8666.527511 | 47928.99 | 1.34 | HARPS+ |
| 8893.696511 | 47930.72 | 1.19 | HARPS+ |
| 8893.867338 | 47925.53 | 0.64 | HARPS+ |
| 8894.613088 | 47928.81 | 0.80 | HARPS+ |
| 8894.776553 | 47927.27 | 0.72 | HARPS+ |
| 8897.551695 | 47931.55 | 0.61 | HARPS+ |
| 8897.735461 | 47930.45 | 0.70 | HARPS+ |
| 8897.874307 | 47930.75 | 0.81 | HARPS+ |
| 8899.651734 | 47928.21 | 0.58 | HARPS+ |
| 8899.827699 | 47928.95 | 0.63 | HARPS+ |
| 8900.608214 | 47927.28 | 0.56 | HARPS+ |
| 8900.789827 | 47928.06 | 0.63 | HARPS+ |
| 8901.564102 | 47924.98 | 0.56 | HARPS+ |
| 8901.747775 | 47926.56 | 0.54 | HARPS+ |
| 8902.715503 | 47928.47 | 0.51 | HARPS+ |
| 8902.892382 | 47926.44 | 0.77 | HARPS+ |

| Time | RV | RV uncertainty | Instrument |
|------|-----|------|------|
| BJD - 2450000 | [m s$^{-1}$] | [m s$^{-1}$] | |
| 8903.665580 | 47928.79 | 0.45 | HARPS+ |
| 8903.831659 | 47927.10 | 0.82 | HARPS+ |
| 8910.570792 | 47928.62 | 0.56 | HARPS+ |
| 8910.756709 | 47929.13 | 0.58 | HARPS+ |
| 8911.555966 | 47927.39 | 0.52 | HARPS+ |
| 8911.706685 | 47927.58 | 0.53 | HARPS+ |
| 8912.683790 | 47924.34 | 0.63 | HARPS+ |
| 8912.844429 | 47922.25 | 0.83 | HARPS+ |
| 8913.640698 | 47925.25 | 0.69 | HARPS+ |
| 8913.809287 | 47921.84 | 0.77 | HARPS+ |
| 8914.613600 | 47927.12 | 0.52 | HARPS+ |
| 8914.767351 | 47924.61 | 1.87 | HARPS+ |
| 8915.576246 | 47923.54 | 0.59 | HARPS+ |
| 8915.758562 | 47921.90 | 0.67 | HARPS+ |
| 8916.545998 | 47927.59 | 0.58 | HARPS+ |
| 8916.712967 | 47925.44 | 0.52 | HARPS+ |
| 8917.531733 | 47924.76 | 0.51 | HARPS+ |
| 8917.691595 | 47924.06 | 0.56 | HARPS+ |
| 8918.642028 | 47925.09 | 0.46 | HARPS+ |
| 8918.797607 | 47926.75 | 0.85 | HARPS+ |
| 8924.680842 | 47924.56 | 0.55 | HARPS+ |
| 8925.668849 | 47925.63 | 0.56 | HARPS+ |
| 8926.686577 | 47925.72 | 0.57 | HARPS+ |
| 8927.755751 | 47928.43 | 0.68 | HARPS+ |
| 8928.651591 | 47927.23 | 0.60 | HARPS+ |
| 8929.692985 | 47928.71 | 0.87 | HARPS+ |
| 8930.729065 | 47931.85 | 0.97 | HARPS+ |
| 8931.713246 | 47930.29 | 0.87 | HARPS+ |

**Table S2. Stellar activity indicators measured from the HARPS spectra.** Columns labelled σ(X) indicate the uncertainty on parameter X. A machine-readable version of the table is available as Data S2.

| Time BJD-2450000 | Hα | σ(Hα) | Hβ | Σ(Hβ) | Hγ | Σ(Hγ) | NaD | σ(NaD) | S | σ(S) |
|---|---|---|---|---|---|---|---|---|---|---|
| 2985.779966 | 0.06352 | 0.00016 | 0.05040 | 0.00037 | 0.1084 | 0.0011 | 0.01141 | 0.00013 | 1.046 | 0.049 |
| 2996.800243 | 0.063398 | 0.000083 | 0.04975 | 0.00019 | 0.11241 | 0.00056 | 0.010365 | 0.000059 | 0.933 | 0.015 |
| 3366.861538 | 0.065270 | 0.000070 | 0.05306 | 0.00016 | 0.11512 | 0.00043 | 0.011170 | 0.000048 | 1.1149 | 0.0099 |
| 3369.821915 | 0.064170 | 0.000065 | 0.05150 | 0.00014 | 0.11359 | 0.00037 | 0.010820 | 0.000044 | 1.0665 | 0.0075 |
| 3370.838755 | 0.065029 | 0.000094 | 0.05302 | 0.00020 | 0.11558 | 0.00053 | 0.011157 | 0.000066 | 1.1194 | 0.0135 |
| 3371.798971 | 0.071996 | 0.000077 | 0.06872 | 0.00018 | 0.15107 | 0.00050 | 0.013621 | 0.000055 | 1.561 | 0.010 |
| 3373.834165 | 0.064966 | 0.000089 | 0.05304 | 0.00020 | 0.11449 | 0.00055 | 0.010926 | 0.000063 | 1.074 | 0.015 |
| 3374.821349 | 0.063813 | 0.000076 | 0.05097 | 0.00017 | 0.11319 | 0.00046 | 0.010721 | 0.000053 | 0.987 | 0.011 |
| 3375.794875 | 0.063731 | 0.000068 | 0.05172 | 0.00015 | 0.11383 | 0.00042 | 0.010668 | 0.000047 | 1.0359 | 0.0088 |
| 3376.803020 | 0.064342 | 0.000076 | 0.05130 | 0.00017 | 0.11515 | 0.00047 | 0.010685 | 0.000052 | 1.039 | 0.011 |
| 3378.725045 | 0.064305 | 0.000070 | 0.05237 | 0.00016 | 0.11334 | 0.00042 | 0.010807 | 0.000048 | 1.0552 | 0.0096 |
| 3491.542478 | 0.06511 | 0.00014 | 0.05238 | 0.00032 | 0.11318 | 0.00087 | 0.01112 | 0.00011 | 1.019 | 0.033 |
| 4120.788138 | 0.064440 | 0.000063 | 0.05228 | 0.00014 | 0.11357 | 0.00035 | 0.010898 | 0.000041 | 1.072 | 0.0061 |
| 4134.749658 | 0.064625 | 0.000080 | 0.05146 | 0.00018 | 0.11328 | 0.00046 | 0.010966 | 0.000054 | 1.060 | 0.011 |
| 4199.598451 | 0.064751 | 0.000083 | 0.05183 | 0.00018 | 0.11373 | 0.00045 | 0.010392 | 0.000053 | 0.9488 | 0.0095 |
| 4231.581759 | 0.064513 | 0.000095 | 0.05137 | 0.00021 | 0.11316 | 0.00057 | 0.010999 | 0.000066 | 1.010 | 0.015 |
| 4257.511359 | 0.065310 | 0.000086 | 0.05214 | 0.00019 | 0.11353 | 0.00049 | 0.010553 | 0.000057 | 1.001 | 0.011 |
| 4569.564214 | 0.064614 | 0.000075 | 0.05052 | 0.00016 | 0.11223 | 0.00043 | 0.010073 | 0.000048 | 0.9540 | 0.0087 |
| 4919.604700 | 0.064050 | 0.000074 | 0.05128 | 0.00016 | 0.11239 | 0.00040 | 0.010374 | 0.000047 | 0.9709 | 0.0075 |
| 4935.546236 | 0.06389 | 0.00011 | 0.05030 | 0.00024 | 0.11226 | 0.00062 | 0.010112 | 0.000075 | 0.914 | 0.016 |
| 4937.573143 | 0.064437 | 0.000092 | 0.05125 | 0.00020 | 0.11243 | 0.00051 | 0.010169 | 0.000059 | 0.938 | 0.011 |
| 4993.478922 | 0.063710 | 0.000096 | 0.04965 | 0.00020 | 0.10993 | 0.00051 | 0.010032 | 0.000061 | 0.926 | 0.012 |
| 5233.697882 | 0.065715 | 0.000096 | 0.05349 | 0.00022 | 0.11397 | 0.00055 | 0.010642 | 0.000063 | 0.996 | 0.013 |

Continued on next page

| Time | Hα | σ(Hα) | Hβ | σ(Hβ) | Hγ | σ(Hγ) | NaD | σ(NaD) | S | σ(S) |
|---|---|---|---|---|---|---|---|---|---|---|
| 5234.675358 | 0.06424 | 0.00014 | 0.05110 | 0.00031 | 0.11060 | 0.00080 | 0.010551 | 0.000096 | 0.905 | 0.025 |
| 8658.453921 | 0.06472 | 0.00011 | 0.05217 | 0.00026 | 0.11412 | 0.00081 | 0.010379 | 0.000071 | 0.813 | 0.029 |
| 8658.461142 | 0.06421 | 0.00011 | 0.05180 | 0.00025 | 0.11147 | 0.00078 | 0.010402 | 0.000070 | 0.8691 | 0.0293 |
| 8658.468654 | 0.06424 | 0.00012 | 0.05180 | 0.00026 | 0.11413 | 0.00084 | 0.010454 | 0.000073 | 0.813 | 0.033 |
| 8658.475609 | 0.06417 | 0.00011 | 0.05257 | 0.00025 | 0.11363 | 0.00078 | 0.010411 | 0.000067 | 0.944 | 0.029 |
| 8658.483201 | 0.06454 | 0.00012 | 0.05315 | 0.00028 | 0.11124 | 0.00088 | 0.010397 | 0.000077 | 0.930 | 0.036 |
| 8658.490238 | 0.06499 | 0.00012 | 0.05417 | 0.00027 | 0.11592 | 0.00086 | 0.010543 | 0.000074 | 0.934 | 0.034 |
| 8658.497610 | 0.06548 | 0.00012 | 0.05403 | 0.00027 | 0.11601 | 0.00088 | 0.010736 | 0.000074 | 0.912 | 0.035 |
| 8658.504844 | 0.06500 | 0.00012 | 0.05374 | 0.00029 | 0.11864 | 0.00092 | 0.010659 | 0.000079 | 0.933 | 0.037 |
| 8658.512158 | 0.06466 | 0.00012 | 0.05302 | 0.00028 | 0.11529 | 0.00091 | 0.010566 | 0.000076 | 0.876 | 0.038 |
| 8658.519681 | 0.06464 | 0.00011 | 0.05272 | 0.00026 | 0.11484 | 0.00085 | 0.010685 | 0.000071 | 0.912 | 0.033 |
| 8658.526845 | 0.06467 | 0.00012 | 0.05252 | 0.00028 | 0.11684 | 0.00092 | 0.010603 | 0.000073 | 0.947 | 0.039 |
| 8658.534217 | 0.06424 | 0.00011 | 0.05196 | 0.00027 | 0.11319 | 0.00088 | 0.010403 | 0.000069 | 0.889 | 0.038 |
| 8658.541242 | 0.06416 | 0.00011 | 0.05180 | 0.00028 | 0.11185 | 0.00093 | 0.010430 | 0.000072 | 0.839 | 0.043 |
| 8665.468148 | 0.06370 | 0.00014 | 0.05071 | 0.00033 | 0.1087 | 0.0011 | 0.010151 | 0.000095 | 0.529 | 0.054 |
| 8665.475520 | 0.06415 | 0.00015 | 0.05114 | 0.00034 | 0.1123 | 0.0011 | 0.010374 | 0.000097 | 0.661 | 0.058 |
| 8665.482893 | 0.06411 | 0.00014 | 0.04907 | 0.00031 | 0.1131 | 0.0011 | 0.010136 | 0.000090 | 0.848 | 0.052 |
| 8665.489223 | 0.06360 | 0.00022 | 0.04962 | 0.00051 | 0.1079 | 0.0017 | 0.01029 | 0.00016 | 0.879 | 0.118 |
| 8665.498748 | 0.06376 | 0.00025 | 0.05134 | 0.00058 | 0.1200 | 0.0020 | 0.01016 | 0.00019 | 0.40 | 0.12 |
| 8665.504107 | 0.06464 | 0.00020 | 0.05415 | 0.00046 | 0.1115 | 0.0015 | 0.01034 | 0.00014 | 0.703 | 0.093 |
| 8665.512937 | 0.06464 | 0.00025 | 0.05231 | 0.00060 | 0.1068 | 0.0021 | 0.01039 | 0.00020 | 0.72 | 0.15 |
| 8665.518712 | 0.06484 | 0.00027 | 0.05145 | 0.00063 | 0.1051 | 0.0022 | 0.01068 | 0.00021 | 0.04 | 0.16 |
| 8665.526779 | 0.06441 | 0.00037 | 0.04787 | 0.00091 | 0.11223 | 0.0037 | 0.01120 | 0.00034 | 0.30 | 0.29 |
| 8665.533526 | 0.06399 | 0.00042 | 0.0543 | 0.0011 | 0.1183 | 0.0046 | 0.01101 | 0.00041 | 0.93 | 0.38 |
| 8665.541107 | 0.0654 | 0.0010 | 0.0607 | 0.0035 | 0.155 | 0.023 | 0.0177 | 0.0013 | -1.12 | 0.91 |
| 8665.548063 | 0.0665 | 0.0015 | 0.0810 | 0.0068 | -0.167 | 0.036 | 0.0171 | 0.0021 | 2.89 | -1.93 |
| 8665.555852 | 0.0646 | 0.0026 | 0.049 | 0.018 | -0.13 | -0.11 | 0.0274 | 0.0041 | -6.19 | 2.48 |
| Continued on next page | | | | | | | | | | |

| Time | Hα | σ(Hα) | Hβ | σ(Hβ) | Hγ | σ(Hγ) | NaD | σ(NaD) | S | σ(S) |
|---|---|---|---|---|---|---|---|---|---|---|
| 8666.447238 | 0.06410 | 0.00010 | 0.05040 | 0.00024 | 0.11361 | 0.00078 | 0.010281 | 0.000066 | 0.711 | 0.029 |
| 8666.454414 | 0.06416 | 0.00012 | 0.05055 | 0.00026 | 0.11195 | 0.00085 | 0.010197 | 0.000076 | 0.798 | 0.034 |
| 8666.461716 | 0.06398 | 0.00012 | 0.05060 | 0.00027 | 0.11196 | 0.00087 | 0.010390 | 0.000077 | 0.781 | 0.036 |
| 8666.468811 | 0.06421 | 0.00014 | 0.05075 | 0.00031 | 0.11341 | 0.00099 | 0.010221 | 0.000088 | 0.759 | 0.044 |
| 8666.476253 | 0.06446 | 0.00014 | 0.05196 | 0.00031 | 0.11156 | 0.00099 | 0.010384 | 0.000090 | 0.878 | 0.044 |
| 8666.483764 | 0.06524 | 0.00016 | 0.05281 | 0.00035 | 0.1168 | 0.0011 | 0.01045 | 0.00010 | 0.749 | 0.052 |
| 8666.491136 | 0.06530 | 0.00015 | 0.05346 | 0.00035 | 0.1161 | 0.0011 | 0.01072 | 0.00010 | 0.874 | 0.052 |
| 8666.498300 | 0.06481 | 0.00015 | 0.05313 | 0.00034 | 0.1105 | 0.0011 | 0.010407 | 0.000099 | 0.757 | 0.051 |
| 8666.505464 | 0.06440 | 0.00017 | 0.05163 | 0.00039 | 0.1076 | 0.0012 | 0.01067 | 0.00012 | 0.632 | 0.065 |
| 8666.512558 | 0.06419 | 0.00018 | 0.05083 | 0.00041 | 0.1190 | 0.0014 | 0.01032 | 0.00013 | 0.776 | 0.073 |
| 8666.520625 | 0.06419 | 0.00019 | 0.05124 | 0.00043 | 0.1029 | 0.0013 | 0.01046 | 0.00013 | 0.799 | 0.078 |
| 8666.527511 | 0.06398 | 0.00018 | 0.05138 | 0.00042 | 0.1077 | 0.0014 | 0.01092 | 0.00013 | 0.711 | 0.077 |
| 8893.696511 | 0.06242 | 0.00017 | 0.04923 | 0.00034 | 0.10962 | 0.00081 | 0.01000 | 0.00013 | 0.779 | 0.030 |
| 8893.867338 | 0.062554 | 0.000081 | 0.05004 | 0.00021 | 0.11052 | 0.00064 | 0.009875 | 0.000055 | 0.782 | 0.032 |
| 8894.613088 | 0.06214 | 0.00011 | 0.04880 | 0.00023 | 0.10864 | 0.00061 | 0.009867 | 0.000074 | 0.783 | 0.022 |
| 8894.776553 | 0.062235 | 0.000098 | 0.04961 | 0.00021 | 0.11054 | 0.00058 | 0.009644 | 0.000064 | 0.769 | 0.022 |
| 8897.551695 | 0.061705 | 0.000082 | 0.04875 | 0.00018 | 0.10973 | 0.00048 | 0.009692 | 0.000052 | 0.784 | 0.016 |
| 8897.735461 | 0.062323 | 0.000096 | 0.04956 | 0.00020 | 0.11160 | 0.00053 | 0.010024 | 0.000063 | 0.819 | 0.018 |
| 8897.874307 | 0.06120 | 0.00010 | 0.04789 | 0.00026 | 0.10950 | 0.00078 | 0.009649 | 0.000073 | 0.757 | 0.041 |
| 8899.651734 | 0.062062 | 0.000081 | 0.04861 | 0.00016 | 0.10901 | 0.00041 | 0.009847 | 0.000052 | 0.783 | 0.013 |
| 8899.827699 | 0.062557 | 0.000083 | 0.04922 | 0.00019 | 0.11137 | 0.00053 | 0.010083 | 0.000057 | 0.785 | 0.023 |
| 8900.608214 | 0.062157 | 0.000076 | 0.04904 | 0.00016 | 0.10863 | 0.00042 | 0.009973 | 0.000050 | 0.796 | 0.013 |
| 8900.789827 | 0.062563 | 0.000082 | 0.04876 | 0.00019 | 0.10944 | 0.00054 | 0.010047 | 0.000057 | 0.782 | 0.025 |
| 8901.564102 | 0.061928 | 0.000077 | 0.04867 | 0.00016 | 0.10909 | 0.00041 | 0.009910 | 0.000050 | 0.815 | 0.013 |
| 8901.747775 | 0.062085 | 0.000072 | 0.04848 | 0.00016 | 0.10886 | 0.00046 | 0.009892 | 0.000048 | 0.819 | 0.019 |
| 8902.715503 | 0.063382 | 0.000071 | 0.05061 | 0.00015 | 0.11122 | 0.00041 | 0.010304 | 0.000046 | 0.862 | 0.014 |
| 8902.892382 | 0.061980 | 0.000094 | 0.04886 | 0.00025 | 0.11110 | 0.00079 | 0.010113 | 0.000072 | 0.832 | 0.047 |

Continued on next page

| Time | Hα | σHα | Hβ | σHβ | Hγ | σHγ | NaD | σNaD | S | σS |
|---|---|---|---|---|---|---|---|---|---|---|
| 8903.665580 | 0.061831 | 0.000062 | 0.04843 | 0.00012 | 0.10827 | 0.00033 | 0.009953 | 0.000039 | 0.8208 | 0.0092 |
| 8903.831659 | 0.06178 | 0.00010 | 0.04905 | 0.00025 | 0.10959 | 0.00074 | 0.009897 | 0.000077 | 0.775 | 0.039 |
| 8910.570792 | 0.062790 | 0.000077 | 0.04892 | 0.00016 | 0.10940 | 0.00041 | 0.009963 | 0.000049 | 0.834 | 0.012 |
| 8910.756709 | 0.062542 | 0.000076 | 0.04950 | 0.00018 | 0.10989 | 0.00052 | 0.010017 | 0.000050 | 0.786 | 0.023 |
| 8911.555966 | 0.062629 | 0.000071 | 0.04913 | 0.00015 | 0.11005 | 0.00040 | 0.009997 | 0.000045 | 0.829 | 0.012 |
| 8911.706685 | 0.063072 | 0.000074 | 0.04985 | 0.00015 | 0.11023 | 0.00041 | 0.010064 | 0.000047 | 0.832 | 0.014 |
| 8912.683790 | 0.062657 | 0.000085 | 0.04839 | 0.00019 | 0.10913 | 0.00052 | 0.009796 | 0.000057 | 0.799 | 0.022 |
| 8912.844429 | 0.06225 | 0.00010 | 0.04846 | 0.00026 | 0.10906 | 0.00081 | 0.009993 | 0.000079 | 0.767 | 0.046 |
| 8913.640698 | 0.062413 | 0.000095 | 0.04907 | 0.00020 | 0.10853 | 0.00053 | 0.009976 | 0.000066 | 0.783 | 0.021 |
| 8913.809287 | 0.063258 | 0.000099 | 0.05205 | 0.00024 | 0.11544 | 0.00074 | 0.010319 | 0.000074 | 0.872 | 0.039 |
| 8914.613600 | 0.062514 | 0.000074 | 0.04932 | 0.00014 | 0.11016 | 0.00040 | 0.010018 | 0.000049 | 0.822 | 0.012 |
| 8914.767351 | 0.06281 | 0.00024 | 0.04915 | 0.00058 | 0.1076 | 0.0022 | 0.010398 | 0.000244 | 0.611 | 0.158 |
| 8915.576246 | 0.062055 | 0.000083 | 0.04893 | 0.00016 | 0.10832 | 0.00048 | 0.009879 | 0.000056 | 0.823 | 0.015 |
| 8915.758562 | 0.062693 | 0.000090 | 0.04954 | 0.00019 | 0.10833 | 0.00063 | 0.009986 | 0.000063 | 0.773 | 0.027 |
| 8916.545998 | 0.062170 | 0.000083 | 0.04839 | 0.00015 | 0.10764 | 0.00046 | 0.009629 | 0.000055 | 0.769 | 0.014 |
| 8916.712967 | 0.062756 | 0.000073 | 0.05051 | 0.00015 | 0.11176 | 0.00048 | 0.009755 | 0.000047 | 0.804 | 0.017 |
| 8917.531733 | 0.062456 | 0.000073 | 0.04966 | 0.00014 | 0.11031 | 0.00042 | 0.009997 | 0.000048 | 0.826 | 0.012 |
| 8917.691595 | 0.062736 | 0.000079 | 0.04934 | 0.00015 | 0.10900 | 0.00050 | 0.009859 | 0.000052 | 0.804 | 0.018 |
| 8918.642028 | 0.063125 | 0.000067 | 0.05019 | 0.00012 | 0.11227 | 0.00038 | 0.010060 | 0.000043 | 0.833 | 0.010 |
| 8918.797607 | 0.06185 | 0.00011 | 0.04889 | 0.00024 | 0.10868 | 0.00081 | 0.009840 | 0.000086 | 0.811 | 0.039 |
| 8924.680842 | 0.063584 | 0.000077 | 0.04966 | 0.00015 | 0.10968 | 0.00049 | 0.009937 | 0.000051 | 0.791 | 0.019 |
| 8925.668849 | 0.064072 | 0.000079 | 0.05100 | 0.00016 | 0.11201 | 0.00050 | 0.010047 | 0.000053 | 0.819 | 0.019 |
| 8926.686577 | 0.063827 | 0.000078 | 0.05020 | 0.00016 | 0.11213 | 0.00051 | 0.009941 | 0.000053 | 0.816 | 0.021 |
| 8927.755751 | 0.063618 | 0.000089 | 0.04914 | 0.00021 | 0.10947 | 0.00070 | 0.009678 | 0.000064 | 0.739 | 0.036 |
| 8928.651591 | 0.063490 | 0.000084 | 0.04977 | 0.00016 | 0.11142 | 0.00049 | 0.009909 | 0.000056 | 0.789 | 0.018 |
| 8929.692985 | 0.06373 | 0.00012 | 0.05040 | 0.00025 | 0.11002 | 0.00075 | 0.009971 | 0.000089 | 0.854 | 0.035 |
| Continued on next page | | | | | | | | | | |

| Time | Hα | σHα | Hβ | σHβ | Hγ | σHγ | NaD | σNaD | S | σS |
|---|---|---|---|---|---|---|---|---|---|---|
| 8930.729065 | 0.06332 | 0.00013 | 0.05074 | 0.00029 | 0.11017 | 0.00094 | 0.00982 | 0.00010 | 0.750 | 0.051 |
| 8931.713246 | 0.06430 | 0.00012 | 0.05062 | 0.00025 | 0.11279 | 0.00080 | 0.010056 | 0.000088 | 0.816 | 0.039 |

**Table S3. Additional stellar data for GJ 367**. Listed are the identifiers of this star in various catalogues, optical and near-infrared photometry, proper motion, and systemic radial velocity. The systemic radial velocity listed is the same as the value quoted in Table 1. The slight discrepancy between these values could have arisen from different instrument zero points, varying systematics present in different instruments, etc.

| Parameter (Unit) | Value | Source |
|---|:---:|:---:|
| **Star: GJ 367b (TOI-731.01)** | | |
| Identifiers | GJ 367, LHS 2182, HIP 47780 | |
| | 2MASS J09442986-4546351 | |
| | TIC 34068865, TOI 731.01 | *(125)* |
| | Gaia DR2 5412250540681250560 | *(8,82)* |
| Photometric magnitudes | | |
| $B$ band | $11.617 \pm 0.034$ | *(125)* |
| $G$ band | $9.1516 \pm 0.0007$ | *(8,82)* |
| $J$ band | $6.632 \pm 0.023$ | *(126)* |
| $H$ band | $6.045 \pm 0.044$ | *(126)* |
| $K$ band | $5.780 \pm 0.020$ | *(126)* |
| Proper motion, right ascension $\mu_{RA}$ (mas yr$^{-1}$) | $-462.549 \pm 0.056$ | *(125)* |
| Proper motion, declination $\mu_{Dec}$ (mas yr$^{-1}$) | $-582.814 \pm 0.058$ | *(125)* |
| Systemic radial velocity*, $V$ (km s$^{-1}$) | $46.96 \pm 0.37$ | *(126)* |
| | $47.740 \pm 0.006$ | *(8,82)* |

**Caption for Data S1:** Machine-readable version of Table S1.

**Caption for Data S2:** Machine-readable version of Table S2.